\documentclass[10pt,preprint]{emulateapj}

\newcommand{\cbo}{B$335$}

\newcommand \msun{\hbox{$\hbox{M}_{\odot}$}}

\begin{document}

\title{\textit{Spitzer} and \textit{HHT} Observations of Bok Globule
B335: Isolated Star Formation Efficiency and Cloud Structure
\altaffilmark{1}}

\author{
Amelia M. Stutz\altaffilmark{2}, 
Mark Rubin\altaffilmark{3},
Michael W. Werner\altaffilmark{3}, 
George H. Rieke\altaffilmark{2},
John H. Bieging\altaffilmark{2}, 
Jocelyn Keene\altaffilmark{4}, 
Miju Kang\altaffilmark{2,5,6}, 
Yancy L. Shirley\altaffilmark{2},
K. Y. L. Su\altaffilmark{2}, 
Thangasamy Velusamy\altaffilmark{3},
David J. Wilner\altaffilmark{7} 
}

\altaffiltext{1}{This work is based in part on observations made with
the \textit{Spitzer Space Telescope}, which is operated by the Jet
Propulsion Laboratory, California Institute of Technology, under NASA
contract 1407.}

\altaffiltext{2}{Department of Astronomy and Steward Observatory,
  University of Arizona, 933 North Cherry Avenue, Tucson, Arizona 85721;
  astutz@as.arizona.edu.}

\altaffiltext{3}{Jet Propulsion Lab, California Institute of
Technology, 4800 Oak Grove Drive, Pasadena, CA 91109.}

\altaffiltext{4}{Caltech, Pasadena, 91125}

\altaffiltext{5}{Korea Astronomy and Space Science Institute, Hwaam
  61--1, Yuseong, Daejeon 305--348, South Korea}

\altaffiltext{6}{Department of Astronomy and Space Science, Chungnam
  National University, Daejeon 305--348, South Korea}

\altaffiltext{7}{Harvard-Smithsonian Center for Astrophysics, 60
  Garden Street, Cambridge, MA 02138}

\begin{abstract}

We present infrared and millimeter observations of Barnard 335, the
prototypical isolated Bok globule with an embedded protostar.  Using
{\it Spitzer} data we measure the source luminosity accurately; we
also constrain the density profile of the innermost globule material
near the protostar using the observation of an 8.0~\micron\ shadow.
{\it HHT} observations of $^{12}$CO 2 -- 1 confirm the detection of a
flattened molecular core with diameter $\sim\!10000$~AU and the same
orientation as the circumstellar disk ($\sim\!100$ to 200~AU in
diameter).  This structure is probably the same as that generating the
8.0~\micron\ shadow and is expected from theoretical simulations of
collapsing embedded protostars.  We estimate the mass of the protostar
to be only $\sim\!5$\% of the mass of the parent globule.

\end{abstract}

\keywords{ISM: globules --  ISM: individual (Barnard 335) --
  infrared: ISM -- stars: formation}

\section{Introduction}

\citet{bok47} and \citet{bok48} were the first to suggest that ``Bok
globules'' could be the sites of isolated star formation. With this in
mind, \cbo\ (CB199 in the catalog of \citet{clemens88}) was
extensively studied in molecular lines \citep{martin78,dickman78} and
in the far-infrared by \citet{keene80} and \citet{keene81}.
\citet{frerking82} discovered a ``pedestal'' source, i.e., a molecular
outflow, indicating the possible presence of a protostar within the
cloud. The presence of a protostar was quickly confirmed with
additional far-infrared observations by \citet{keene83}.  Assuming a
distance of 250 pc, as suggested by \citet{tomita79} and as adopted by
most subsequent observers, \citet{keene83} determined the luminosity
of the source within \cbo\ to be about 3 L$_{\odot}$.  \cbo\ has
subsequently been observed at far-infrared and submillimeter
wavelengths by, e.g., \citet{gee85}, \citet{mozu86},
\citet{chandler90}, \citet{huard99}, and \citet{shirley00}; it is the
first example of a Class 0 young stellar object to be so thoroughly
studied.  It is therefore the template source for many of our concepts
about the early stages of the formation of low-mass stars.

Modeling has established that the globule material is still undergoing
strong noncircular motions that are most likely infall
\citep[e.g.,][]{zhou90,choi95,velusamy95}.  Although this scenario of
infall, and more specifically of inside-out collapse, is plausible,
important discrepancies remain
\citep[e.g.,][]{wilner00,harvey01,evans05,choi07}.  Models and
observations imply that the core is a star of mass $\sim\!0.5$~\msun
\citep{zhou90} with a circumstellar disk $\sim\!100$ to 200~AU in
diameter \citep{harvey03a}.  This source has a strong bipolar outflow
\citep[e.g.,][]{cabrit92,hirano88}, oriented nearly in the plane of
the sky, that maintains a cavity in the interstellar medium (ISM) and
is associated with the observed shocked material \citep{galfalk07}.

We describe observations with all three instruments on {\it Spitzer}
and with the Heinrich Hertz Submillimeter Telescope that allow us to
1) measure the density distribution in the center of the globule; 2)
constrain the geometry of the outflow and thus the protostellar disk;
and 3) measure the source luminosity accurately.  These measurements
(with a revised distance of 150~pc) reveal a flattened molecular
cloud core, consistent with numerical simulations of cloud collapse;
provide tight constraints on the opening angle of the outflow; and
demonstrate internal consistency between the globule luminosity (and
other characteristics) and those expected for a 0.3 to 0.4~\msun\
young star.

\begin{figure*}[t]
  \scalebox{1.15}{\plotone{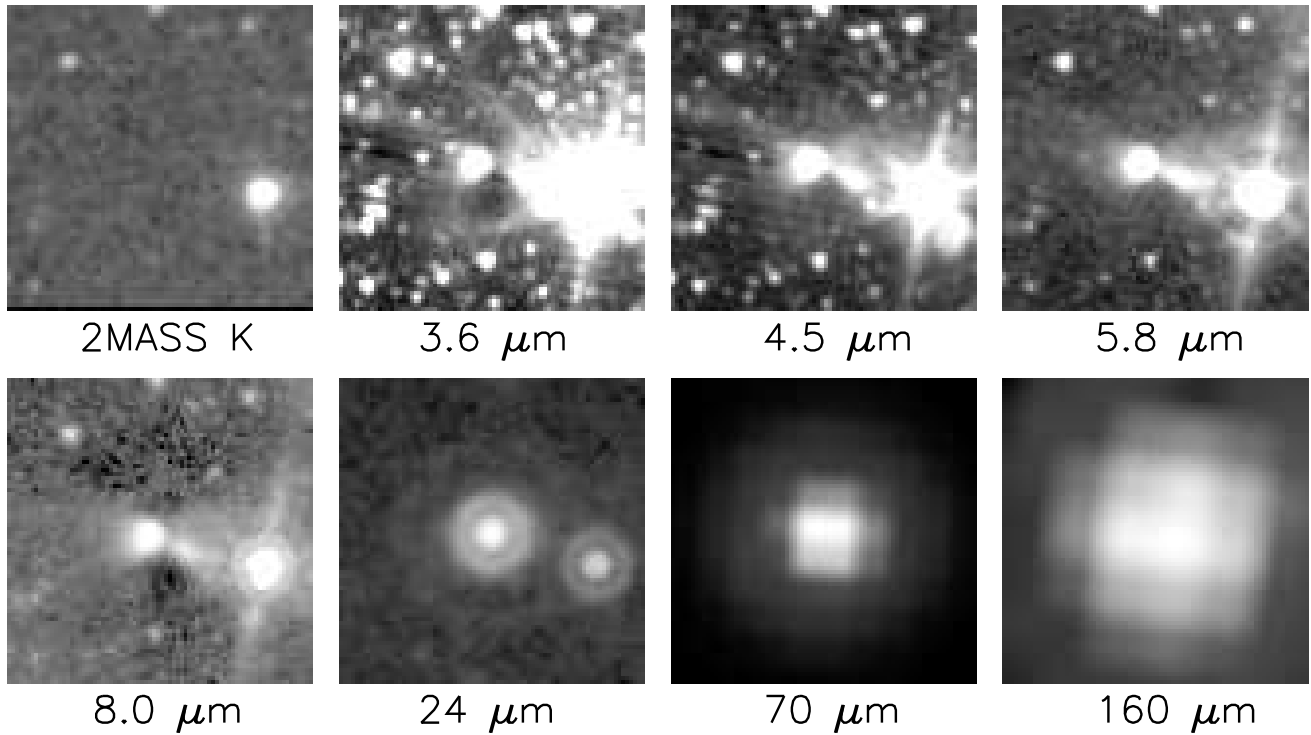}}
  \caption{Gallery of images of B335 (CB199) at the indicated
  wavelengths.  Images are shown with a logarithmic intensity scale
  and are 1$\farcm$6 on a side.  North is up and east is to the left.}
  \label{fig:mips}
\end{figure*}

\section{Observations and processing}

\subsection{Spitzer data}

\cbo\ was observed by the {\it Spitzer Space Telescope} on 2004 April
20 and 21 with the Infrared Array Camera \citep[IRAC;][]{fazio04}
AORKEYs 4926208 and 4926464; 2004 October 15 with the Multiband
Imaging Photometer \citep[MIPS;][]{rieke04} AORKEY 12022106; and 2004
October 14 with the Infrared Spectrograph \citep[IRS;][]{houck04}
AORKEY 3567360.  Observations were made at two closely separated
epochs to permit the elimination of asteroids from the final mosaic.
The field of view with coverage in all four IRAC bands is $5.5\arcmin
\times 13.6\arcmin$, with a larger field covered in at least two
bands.  For each IRAC band, 3 overlapping positions arranged along a
column were imaged.  At each position five small Gaussian dithers of
10.4~s duration were taken.  In the region where we had coverage in
all four bands, this resulted in an average exposure time of about
120~sec.  For each dither, the 10.4 second long-integration frame was
accompanied by a short-integration frame of 0.4s in ``high dynamic
range (HDR)'' mode.  This allowed for correction of saturated pixels
in the longer exposures.

\begin{figure}[t]
  \begin{center}
    \scalebox{0.49}{{\includegraphics{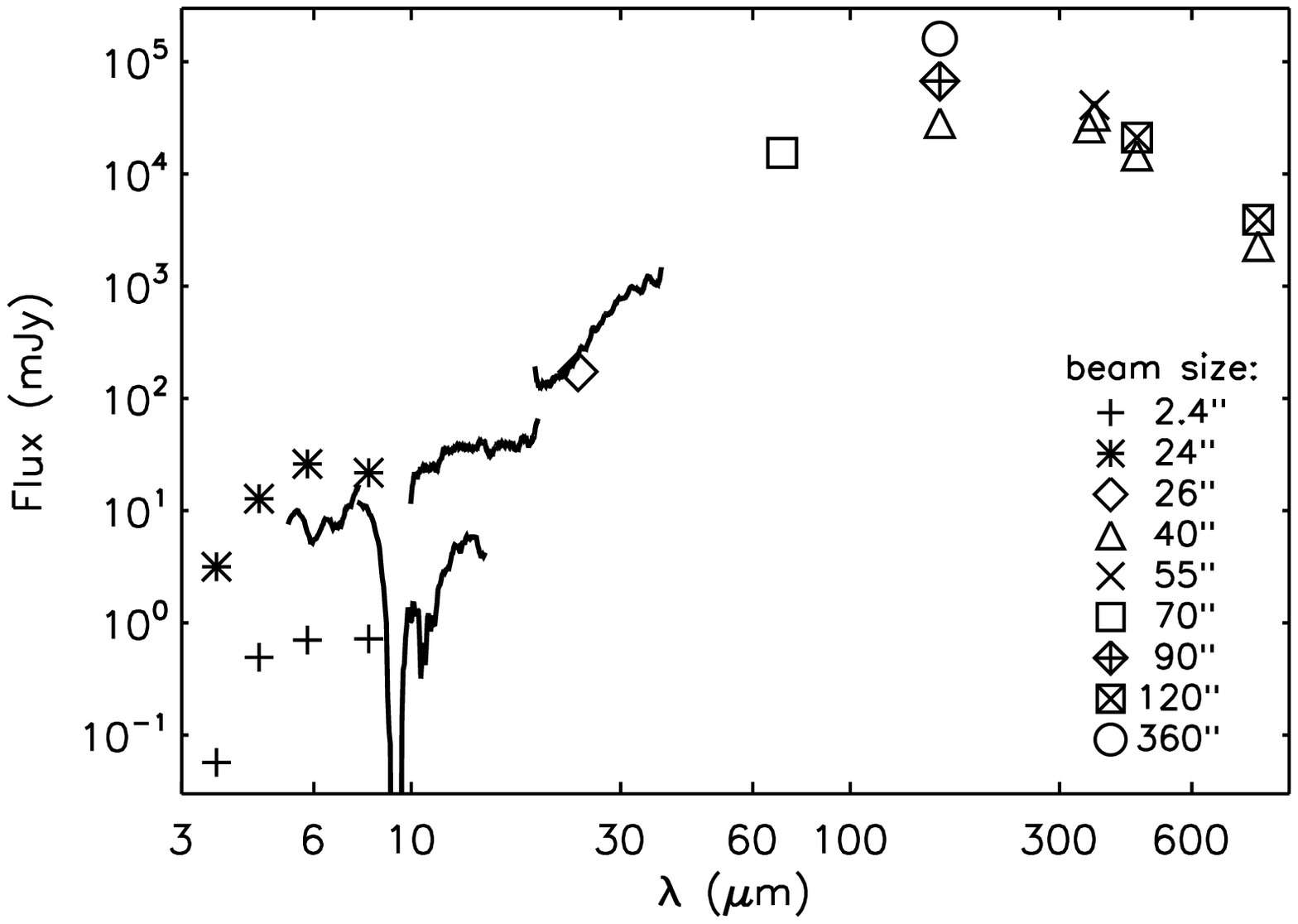}}}
    \scalebox{0.49}{{\includegraphics{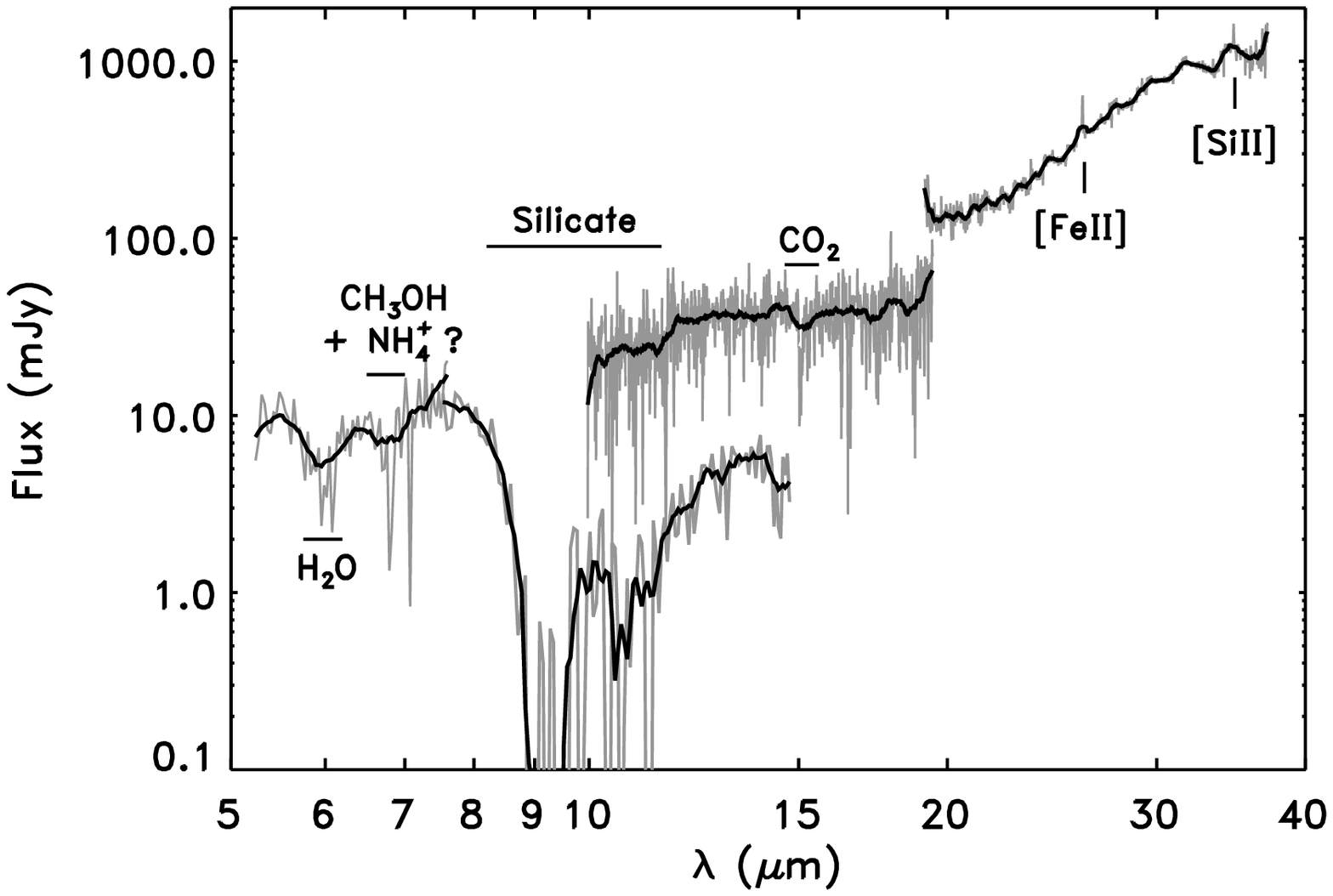}}}
    \caption{\small{SED and spectrum of \cbo.  {\it Top panel:}
      Different symbols represent fluxes measured at the different
      beam sizes indicated in the legend; these data are listed in
      Table~2.  The magnitude of possible aperture corrections in the
      3 to 10~\micron\ region can be judged by the crosses, which are
      photometry measured in a $1\farcs2$ aperture (point-source).
      The black line shows the IRS data smoothed to
      $\sim\!0.4$~\micron.  {\it Bottom panel:} The grey line shows
      the IRS spectrum at full resolution; the black line shows the
      same data smoothed to $\sim\!0.4$~\micron.  Various features are
      indicated in the spectrum.  {\it Both panels:} The IRS spectra
      do not join because the high resolution spectrum, from 10 to
      20~\micron, has a larger slit than the low resolution spectrum,
      from 5 to 14~\micron; hence the longer wavelengths include more
      extended emission.  }}
    \label{fig:irs}  
  \end{center}
\end{figure}

MIPS instrument observations were carried out using scan map mode in
all three MIPS bands at 24~\micron, 70~\micron, and 160~\micron\ as
part of Spitzer program ID 53 (P.I. G. Rieke).  The data reduction was
carried out using the Data Analysis Tool \citep[DAT;][]{gordon05} out
following the steps outlined in a paper by \citet{stutz07} describing a
similar study of the Bok Globule CB190; please refer to that
publication for details.  For \cbo, the observations covered a field
of $\sim\!15\arcmin \times 55\arcmin$, with an exposure time of
$\sim\!192$~s at 24~\micron, $\sim\!80$~s at 70~\micron, and
$\sim\!20$~s at 160~\micron, although the integration time varies
significantly in each map due to non--uniform coverage.

\begin{figure*}
  \begin{center}
    \scalebox{0.91}{{\includegraphics[angle=-90]{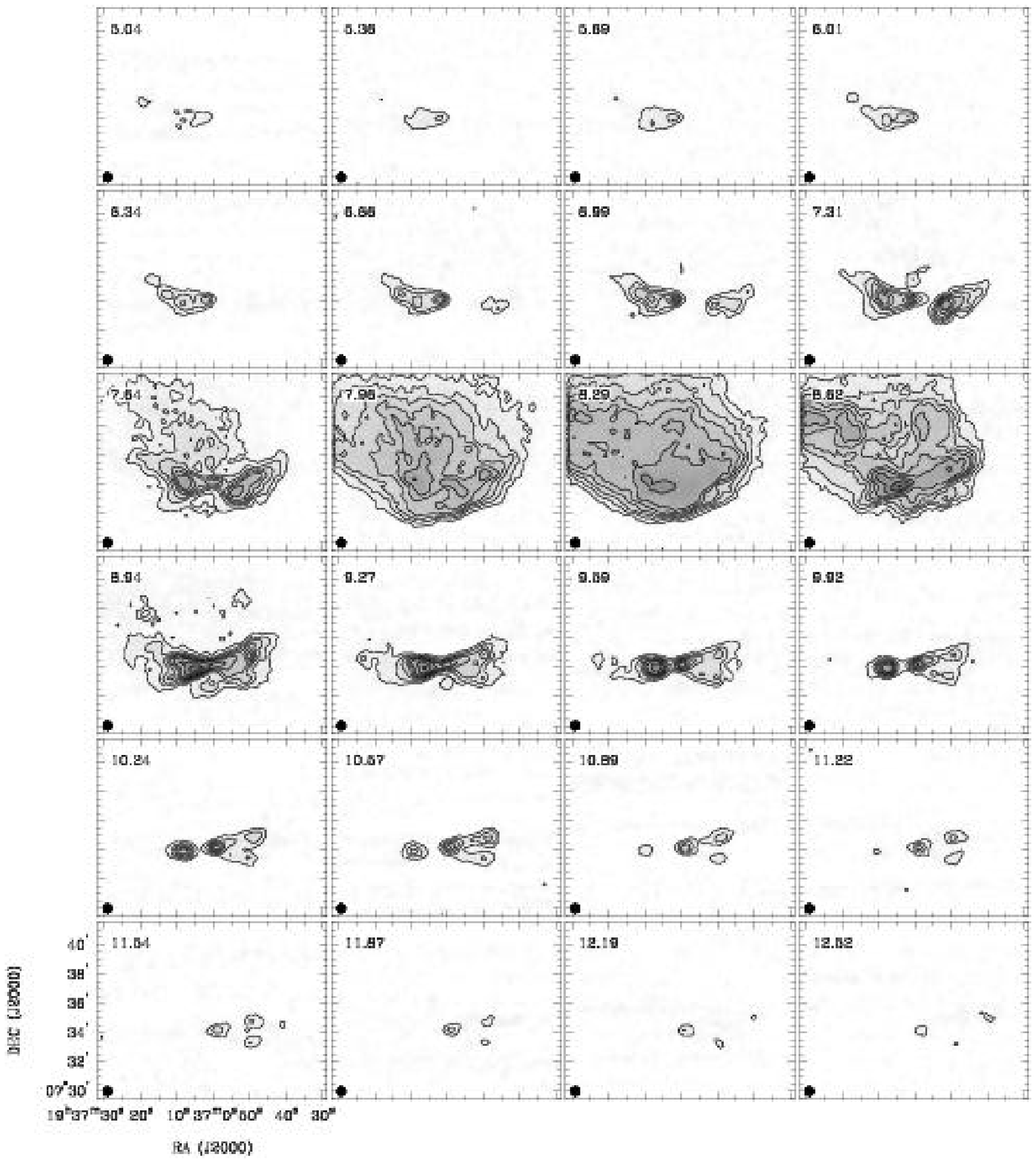}}}
    \caption{Large $^{12}$CO J = 2 -- 1 OTF channel map of \cbo,
    $16\arcmin \times 12\arcmin$ in size, showing channels with
    significant emission, from 12.52~km~s$^{-1}$ to 5.04~km~s$^{-1}$,
    at a $\sim\!0.3$~km~s$^{-1}$ resolution.  Channel velocities are
    indicated in the top left corner and the beam size ($32\arcsec$)
    is indicated as the filled circle in the bottom left of each
    panel.  Contour levels are $\{1,2,3,4,5,6,8,10,12,14,16\} \times
    0.5$~K-T$_{\rm A}^*$.}
    \label{fig:12co}
  \end{center}
\end{figure*} 

\begin{deluxetable*}{lcccccc}[t]
\tabletypesize{\scriptsize}
\tablecaption{Summary of Outflow Parameters}
\tablewidth{0pt}
\tablehead{
\colhead{Dist.} 
& \colhead{M$_{\rm thin}$} 
& \colhead{R$_{\rm CO}$}
& \colhead{V$_{\rm CO}$} 
& \colhead{T$_{\rm CO}$} 
& \colhead{F$_{\rm obs}$} 
& \colhead{L$_{\rm obs}$} \\
\colhead{[pc]}
& \colhead{[$\msun$]}
& \colhead{[pc]}
& \colhead{[km s$^{-1}$]}
& \colhead{[yr]}
& \colhead{[$\msun$ km s$^{-1}$ yr$^{-1}$]}
& \colhead{[$\msun$ cm$^2$ s$^{-2}$ yr$^{-1}$]}\\
\colhead{(1)}
& \colhead{(2)} 
& \colhead{(3)} 
& \colhead{(4)} 
& \colhead{(5)} 
& \colhead{(6)} 
& \colhead{(7)} 
}
\startdata
150 &  0.008 & 0.21 & 4.55 & 4.6$\times$10$^4$ & 7.9$\times$10$^{-7}$ & 1.8$\times$10$^4$ \\
250 &  0.02  & 0.36 & 4.55 & 7.7$\times$10$^4$ & 1.3$\times$10$^{-6}$ & 3.0$\times$10$^4$ \\
\enddata
\label{tab:co}
\tablecomments{ Outflow parameters are calculated according to the
  procedure described in \citet{cabrit92} and are derived from the
  large ($16\arcmin \times 12\arcmin)$ $^{12}$CO and $^{13}$CO maps
  presented in Figures~\ref{fig:12co} and \ref{fig:13co}. }
\end{deluxetable*}
                                                                                    
\begin{figure}
  \begin{center}
    \scalebox{0.42}{{\includegraphics{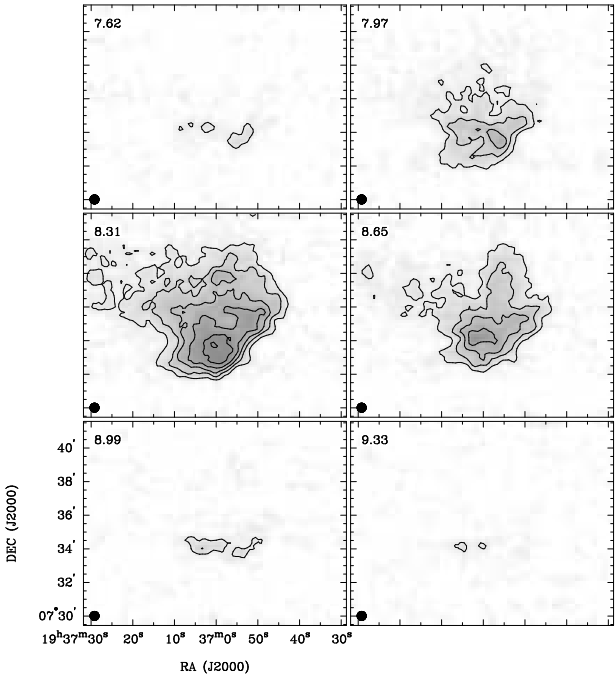}}}
    \caption{Large $^{13}$CO J = 2 -- 1 OTF channel map of \cbo, same
     size and spectral resolution as Figure~\ref{fig:12co}.  Channels
     with significant emission are shown, from 7.62~km~s$^{-1}$ to
     9.33~km~s$^{-1}$.  Channel velocities are indicated in the top
     left corner and the beam size ($32\arcsec$) is indicated as the
     filled circle in the bottom left of each panel.  Contour levels
     are $\{1,2,3,4,5,6\} \times 0.5$~K-T$_{\rm A}^*$.}
    \label{fig:13co}
  \end{center}
\end{figure}

\begin{deluxetable}{crrr} 
\tabletypesize{\scriptsize}
\tablecaption{\cbo\ Photometry}
\tablewidth{0pt}
\tablehead{
\colhead{$\lambda$}
&\colhead{Beam}
&\colhead{Flux}
&\colhead{Error}\\
\colhead{[$\micron$]}
&\colhead{[$\arcsec$]}
&\colhead{[mJy]}
&\colhead{[mJy]}\\
\colhead{(1)}
& \colhead{(2)} 
& \colhead{(3)} 
& \colhead{(4)} 
}
\startdata
3.6       &    2.4   &    0.06  &    0.01  \\
``        &   24.0   &    3.16  &    0.30  \\
4.5       &    2.4   &    0.49  &    0.05  \\
``        &   24.0   &   12.76  &    1.00  \\
5.8       &    2.4   &    0.70  &    0.07  \\
``        &   24.0   &   26.06  &    2.50  \\
8.0       &    2.4   &    0.72  &    0.07  \\
``        &   24.0   &   21.77  &    2.00  \\ 
24        &   26.0   &   173    &   16     \\
70        &   70.0   & 15415    & 2090     \\
160       &   40.0   & 28397    &  6167    \\
``        &   90.0   & 67250    & 15250    \\
``        &  100.0   & 68091    & 15231    \\
``        &  360.0   & 160360   & 31090    \\
350$^a$   &   40.0   & 25900    &  4100    \\ 
360$^b$   &   40.0   & 33000    &  1700    \\
360$^b$   &   55.0   & 41000    &  2800    \\
450$^c$   &   40.0   & 14600    &  2200    \\
450$^c$   &  120.0   & 21100    &  3300    \\
850$^c$   &   40.0   &  2280    &   300    \\
850$^c$   &  120.0   &  3910    &   220    \\

\enddata
\label{tab:phot}
\tablenotetext{a}{\citet{wu07}}
\tablenotetext{b}{\citet{gee85}}
\tablenotetext{c}{\citet{shirley00}}
\end{deluxetable}

The IRAC data were processed by the Spitzer Science Center (SSC) using
the standard data pipeline (version 11.4) to produce basic calibrated
data (BCD) images.  To process the BCDs and produce mosaics, we used a
set of IDL procedures and csh scripts (for data handling and module
control) that were provided by Sean Carey of the Spitzer instrument
support team.  These routines use existing MOPEX software developed by
the Spitzer Science Center \citep{makovoz05}.  We produced final
mosaics with a pixel scale of $0\farcs6$ (see Figure~\ref{fig:mips}).
Source extraction and photometry were performed using the SExtractor
software \citep{bertin96}.  An aperture with a diameter of 8 pixels
was used in the extraction.  Aperture correction factors, derived by
the Spitzer Science Center and provided in the IRAC Data Handbook 2.0,
were used to estimate the true flux of objects based on the measured
flux in each aperture.  These fluxes were then converted into
magnitudes using the zero magnitude flux values taken from the IRAC
Data Handbook.  The overall flux calibration, expressed in physical
units of MJy sr$^{-1}$, is estimated by the SSC to be accurate to
within 10\%.  All the Spitzer images are displayed in
Figure~\ref{fig:mips}; even at 8.0~\micron\ the embeded point source
is not visible and the images are dominated by reflection nebulosity.
The Spitzer photometry is plotted in Figure~\ref{fig:irs} along with
longer wavelength data taken from the literature.

IRS observations of B335 were taken in mapping mode as part of the IRS
Disks GTO program 02, AOR key 3567360.  The source was observed with
three modules: short-hi, long-hi, and short-lo [both first and second
orders].  A small raster map consisting of two positions parallel to
the slit and three perpendicular to the slit was made with each
module.  The raster spacings parallel to the slit were selected so
that the usual IRS staring mode point source measurement - which
consists of measurements at positions one third of the slit length
from either end - could be reconstructed from the central of the three
perpendicular positions.  The spacings perpendicular to the slit were a
little more than half of the slit width.  The observations at each
raster position were carried out with an integration time of
6.3 sec.  

Examination of the two dimensional spectral images shows the source to
be compact along the slit.  Thus, the reduction has been carried out
using only the central perpendicular position from each module, as
though the data were the result of an IRS staring mode observation.
The position of this point source, as defined by the raster
parameters, is RA = $19^h 37^m 0.89^s$, Dec = $+07^o 34\arcmin
9.84\arcsec$ (all positions in this work are given in the J2000
system).  A small amount of extended emission may have been missed in
this process.  The synthetic point source observations were reduced
using the Cubism software \citep{smith07}.  The spectrum is shown in
Figure~\ref{fig:irs}.

\subsection{$^{12}$CO and $^{13}$CO data}

The \cbo\ region was mapped in the J=2--1 transitions of $^{12}$CO and
$^{13}$CO with the 10-m diameter Heinrich Hertz Telescope (HHT) on
Mt. Graham, Arizona on April 22nd, 2007.  The observations were made
with the 1.3mm ALMA sideband separating receiver with a 4 - 6 GHz IF
band.  The $^{12}$CO J=2-1 line at 230.538 GHz was placed in the upper
sideband and the $^{13}$CO J=2-1 line at 220.399 GHz in the lower
sideband, with a small offset in frequency to ensure that the two
lines were adequately separated in the IF band.  The spectrometers,
one for each of the two polarizations, were filter banks with 256
channels of 250 KHz width and separation.  At the observing
frequencies, the spectral resolution was $\sim\!0.3$~km~s$^{-1}$ and
the angular resolution of the telescope was 32$\arcsec$ (FWHM).  The
main beam efficiency, measured from planets, was $0.85\pm0.4$.

A $16\arcmin \times 12\arcmin$ field centered at RA = $19^h 37^m
00.4^s$, Dec = $+07^o 35\arcmin 29\arcsec$ was mapped with on-the-fly
(OTF) scanning in RA at $10\arcsec$~s$^{-1}$, with row spacing of
$10\arcsec$ in declination, over a total of 72 rows.  System
temperatures were calibrated by the standard ambient temperature load
method \citep{kutner81} after every other row of the map grid.
Atmospheric conditions were clear and stable, and the system
temperatures were nearly constant, and averaged $T_{sys} = 191$~K
(SSB).

Data for each CO isotopomer were processed with the {\it CLASS}
reduction package (from the University of Grenoble Astrophysics
Group), by removing a linear baseline and convolving the data to a
square grid with $10\arcsec$ grid spacing (equal to one-third the
telescope beamwidth).  The intensity scales for the two polarizations
were determined from observations of standard sources made just before
the OTF maps.  The gridded spectral data cubes were processed with the
{\it Miriad} software package \citep{sault95} for further analysis.
This process yielded images with rms noise per pixel and per velocity
channel of 0.21~K-T$_A ^*$ for both the $^{12}$CO and $^{13}$CO
transitions.  The $^{12}$CO and $^{13}$CO channel maps are shown in
Figures~\ref{fig:12co} and~\ref{fig:13co}, respectively.

\begin{figure*}
  \begin{center}
    \scalebox{0.9}{{\includegraphics{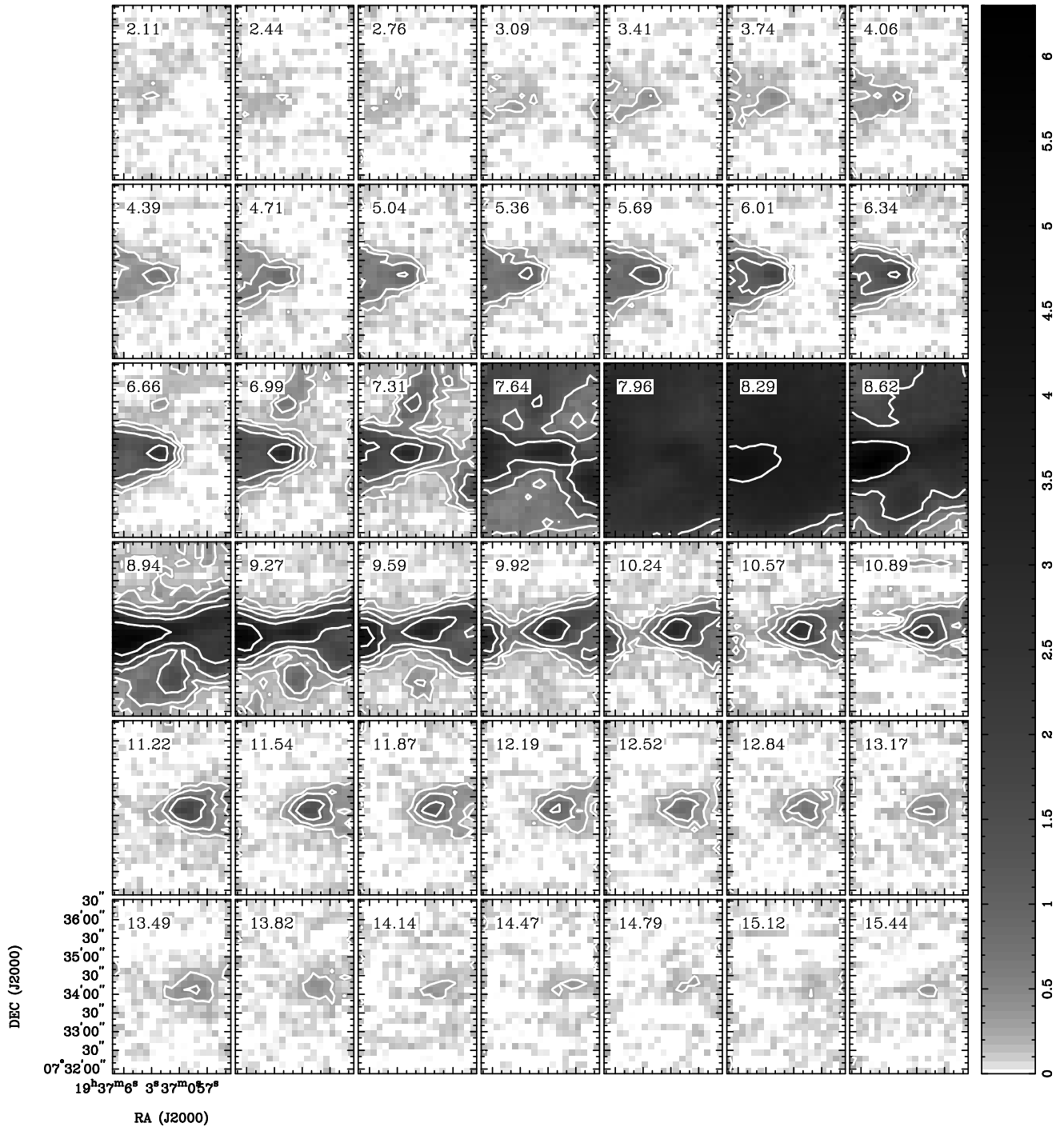}}}
    \caption{$^{12}$CO J = 2 -- 1 OTF sub-field channel map of \cbo,
      $2\farcm5 \times 4\farcm5$ in size, centered on RA = $19^h 37^m
      00.8^s$, Dec = $+07\degr 34\arcmin 07\farcs4$, near the location
      of the protostar.  Channels with significant emission are shown,
      from 15.44~km~s$^{-1}$ to 2.11~km~s$^{-1}$, observed at a
      resolution $\sim\!0,3$~km~s$^{-1}$.  The temperature scale is
      indicated in the right side; contour levels are $\{1,2,4,8,16\}
      \times 0.25$~K-T$_{\rm A}^*$.  Note the structures located north
      and south of the main east-west outflow region, visible here in
      the south at $\sim\!9.27$~km~s$^{-1}$ and in the north at
      $\sim\!7.31$~km~s$^{-1}$.}
    \label{fig:12cosub}
  \end{center}
\end{figure*}

Using the same set-up as that outlined above, we made a deeper map of
a $2\farcm5 \times 4\farcm5$ region centered at RA = $19^h 37^m
00.8^s$, Dec = $+07^o 34\arcmin 07\arcsec$, near the protostar.  These
observations were carried out on March 19, 2008, and were repeated 7
times.  The average $T_{sys} = 205$~K (SSB), and was nearly constant
throughout the observations.  After combining the 14 images with
weights proportional to the individual map rms values, we obtain final
rms values per pixel and per velocity channel of $\sigma_{\rm T_{\rm
A}^*} = 0.091$~K ($^{12}$CO) and $\sigma_{\rm T_{\rm A}^*} =0.084$~K
($^{13}$CO).  These maps are shown in Figures~\ref{fig:12cosub}
and~\ref{fig:13cosub}, while Figure~\ref{fig:pv} shows
position-velocity diagrams derived from the same data.

\section{The Distance}

A quantitative interpretation of our observations requires an estimate
of the distance to \cbo.  Here we examine the often-quoted and widely
accepted \citet{tomita79} distance to \cbo\ of 250~pc and update it
with modern data.

\citet{tomita79} derive a distance to \cbo\ ranging from 130 to
250~pc, based on photoelectric stellar data from the \citet{blanco70}
catalog, and a distance of 600 pc based on star counts.  They adopt
250~pc as a plausible value but do not provide further details or show
figures of $A_V$ vs.\ distance or of stellar density in the field of
interest.  The \citet{blanco70} catalog data in the region of \cbo\
originates from \citet{bouigue61} catalog photometry.  For
completeness we note that the origin of the spectral types of the
stars listed in the \citet{bouigue61} catalog are from an evidently
unpublished provisional spectral classification determined by
l'Observatoire de Marseille; the quoted spectral types appear
generally reliable but do not always agree with those quoted in
\textit{SIMBAD\footnote{http://simbad.u-strasbg.fr/simbad/}}.  The
\citet{blanco70} catalog references 9 stars \citep{bouigue61} within
45$\arcmin$ of \cbo, well outside the optical extinction extent of
\cbo, $r \sim 3\arcmin$.  Therefore, the \citet{tomita79} distance of
250~pc is highly uncertain.

We show an updated version of the $A_B$ vs.\ distance plot in
Figure~\ref{fig:dist}.  The data are from the on-line
\textit{VizieR\footnote{http://vizier.u-strasbg.fr/viz-bin/VizieR}}
service.  We find 32 stars within $20\arcmin$ of \cbo\ with spectral
classifications.  Of these, we reject all stars without both B-band
and K-band photometry, as well as those with spectral classifications
of G2 or later.  Later-type stars mostly add scatter to our analysis
because a small error in the spectral type yields a large error in
both $A_B$ and distance.  In all cases for which we do not have a
luminosity class we assume that the star is on the main sequence.  We
assume $A_B = 1.09 \times E(B - K)$ \citep{rieke85}.  The resulting
data are presented in fig.~\ref{fig:dist}.  We also indicate the
\citet{tomita79} distance estimates (at 130, 250, and 600~pc) as
dotted lines.  We do not see any noticeable feature at or near 250~pc.
To put a lower bound on the distance we use the \citet{lallement03}
maps of the Local Bubble; in the direction of \cbo\ ($l = 44.9^o, b =
-6.6^o$) they measure the extent of the local low density bubble to be
$\sim\!60$~pc.  We set the upper limit to the distance to \cbo\ to be
$\sim\!200$~pc since in Figure~\ref{fig:dist} we see approximately
constant stellar $A_B$ values outside of this distance.  We adopt a
distance of 150~pc in the following discussion, indicated as a solid
line in Figure~\ref{fig:dist}.  Our adopted distance is uncertain by
nearly a factor of two, but it falls near the middle of the plausible
range.  It also reduces a number of discrepancies found by
\citet{harvey01} in modeling the extinction of the globule, and helps
reconcile the inferred properties for the protostar with the
luminosity of the globule (see \S~5).

\section{Resolved Structure}

\subsection{The Outflow}

\begin{figure}[t]
  \begin{center}
    \scalebox{0.32}{{\includegraphics[angle=-90]{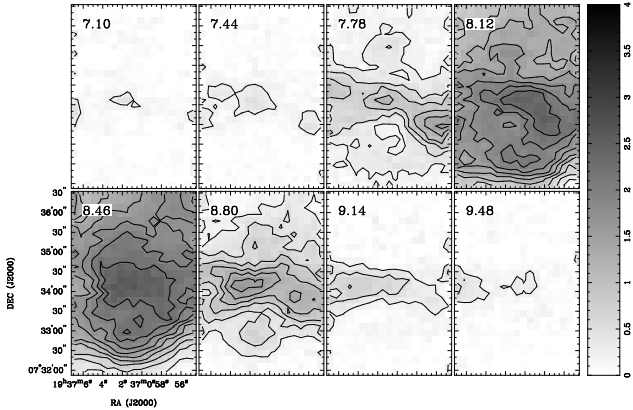}}}
    \caption{$^{13}$CO J = 2 -- 1 map; same as
    Figure~\ref{fig:12cosub}.  Channels with significant emission are
    shown, from 7.10~km~s$^{-1}$ to 9.48~km~s$^{-1}$.  Contour levels
    are $\{1,2,3,4,5,6,7,8,9\} \times 0.25$~K-T$_{\rm A}^*$. }
    \label{fig:13cosub}
  \end{center}
\end{figure}

\begin{figure}
  \begin{center}
    \scalebox{0.29}{{\includegraphics{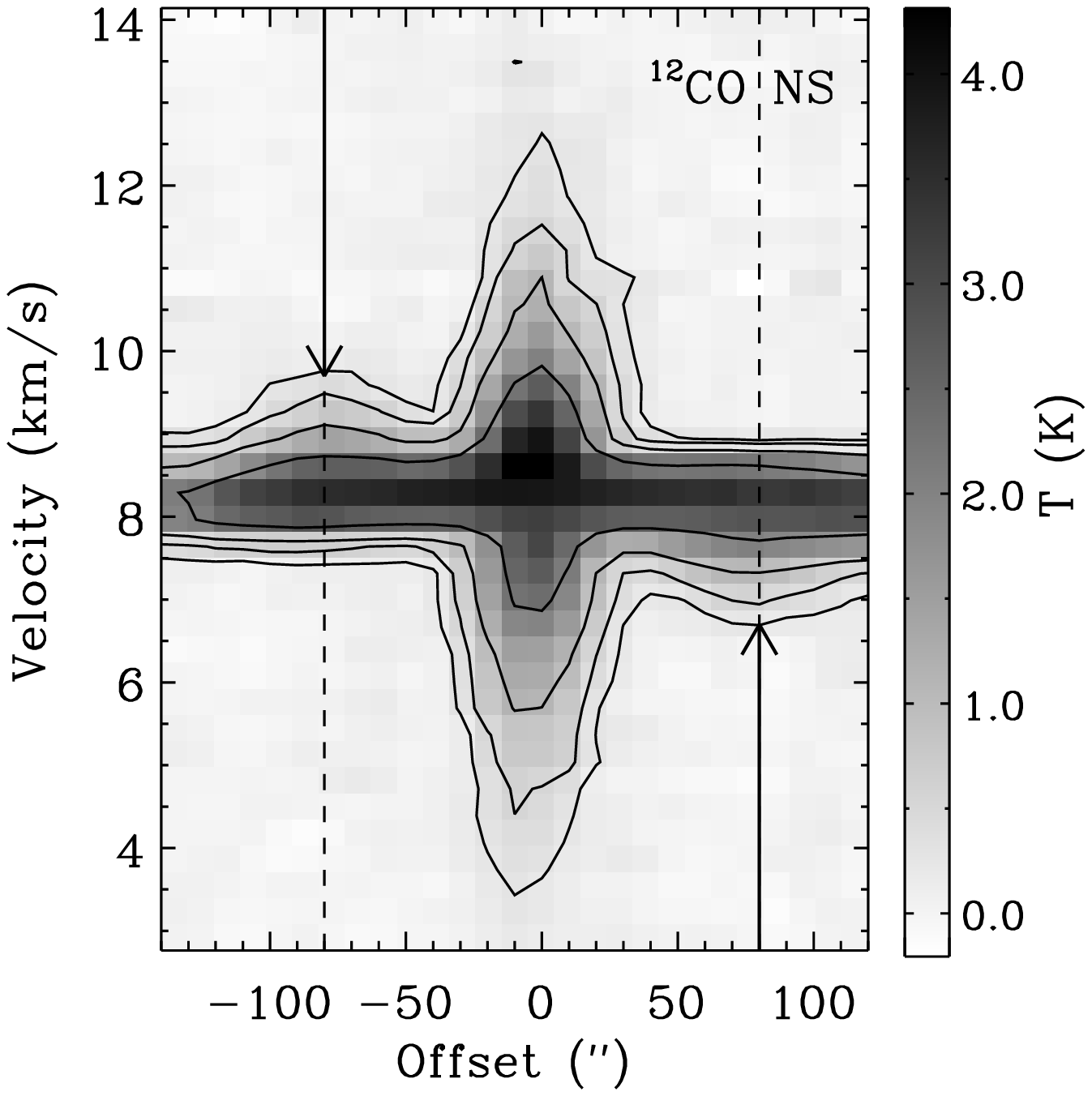}}}
    \scalebox{0.29}{{\includegraphics{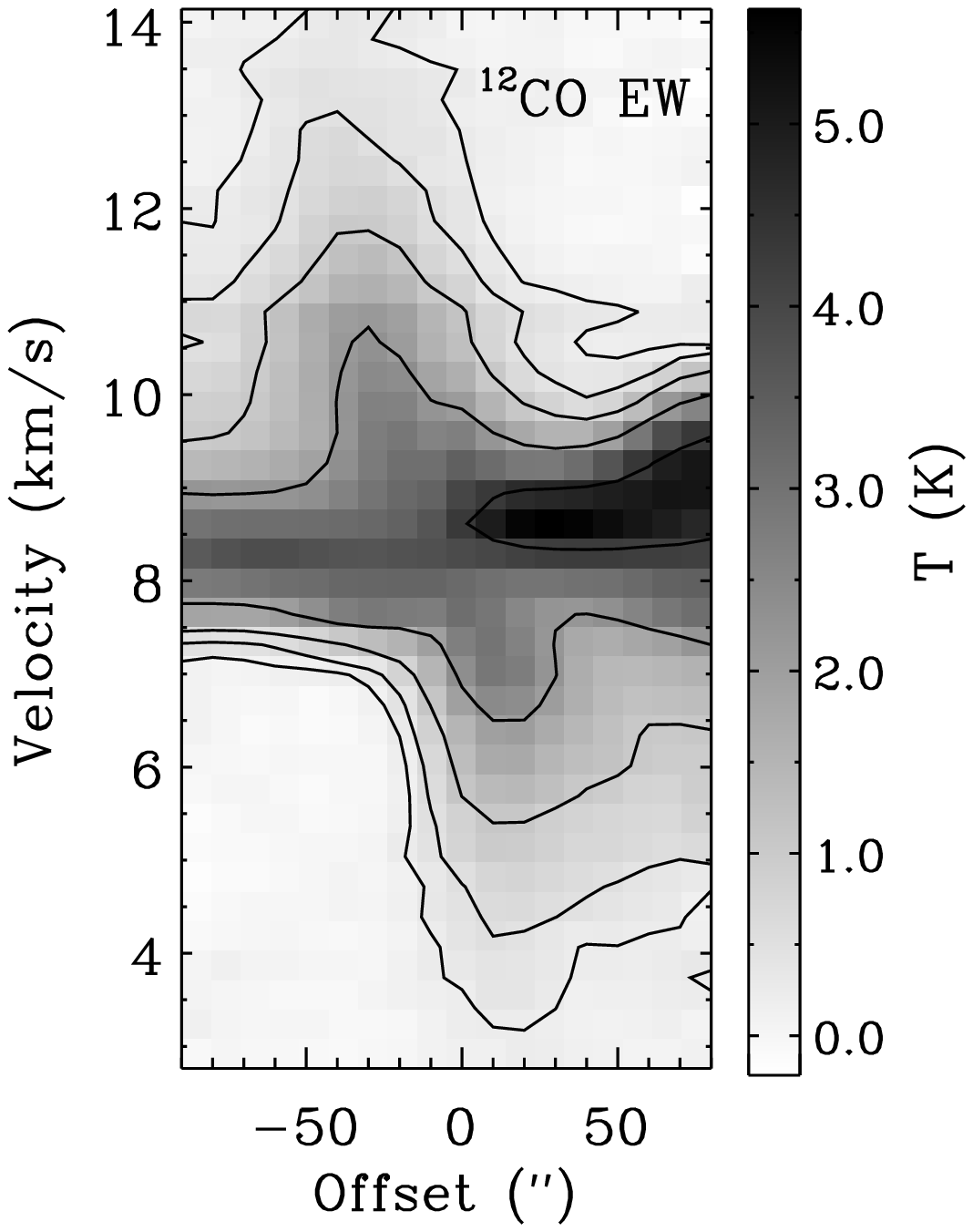}}}
    \scalebox{0.29}{{\includegraphics{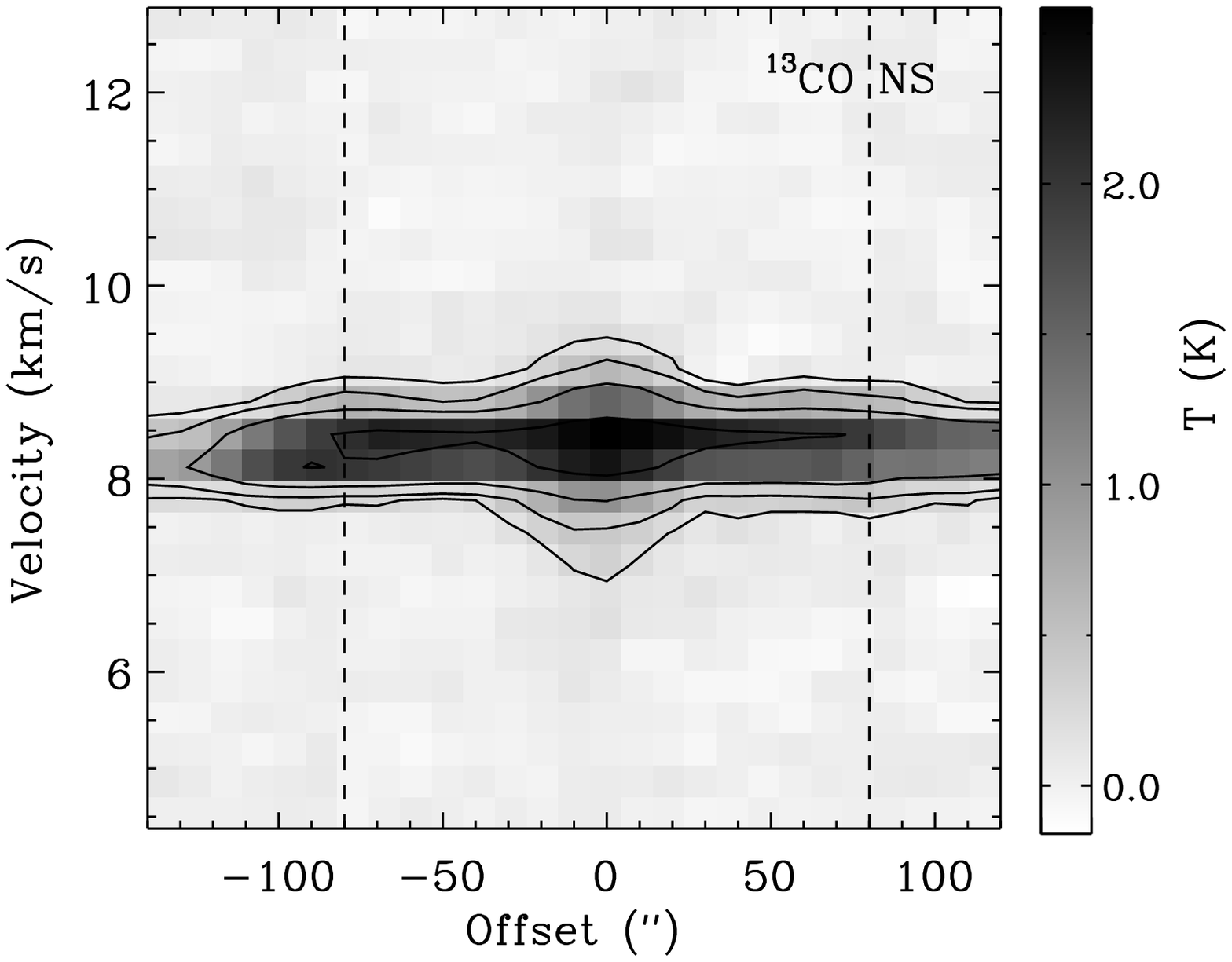}}{\includegraphics{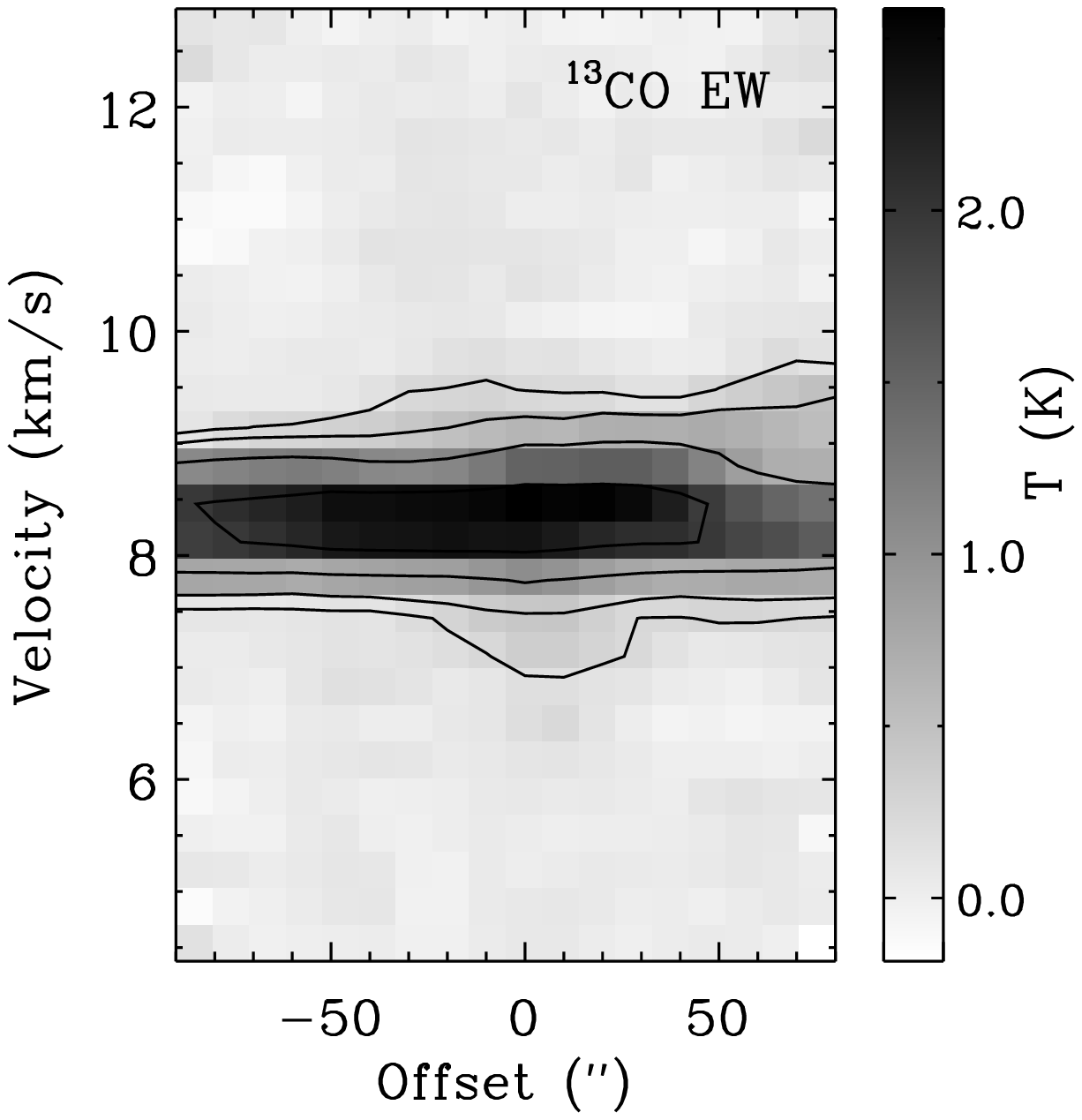}}}
    \caption{$^{12}$CO ({\it top}) and $^{13}$CO ({\it bottom})
     position-velocity diagrams for the sub--field map.  The $^{12}$CO
     contour levels are $\{1,2,4,8,16\} \times 3\sigma$, where
     $3\sigma = $0.28~K-T$_{\rm A}^*$.  The $^{13}$CO contour levels
     are also $\{1,2,4,8\} \times 3\sigma$, where $3\sigma =
     0.25$~K-T$_{\rm A}^*$.  The $^{12}$CO north-south cut is centered
     on RA = $19^h 37^m  0.75^s$; the east-west cut is centered on Dec
     = $07^o 34\arcmin 7\farcs05$.  The $^{13}$CO north-south cut is
     centered on RA = $19^h 37^m 0.83^s$; the east-west cut is
     centered on Dec = $07^o 34\arcmin 5\farcs26$. The vertical lines
     (dashed and solid arrows) in the $^{12}$CO and $^{13}$CO NS cuts
     indicate the offsets ($80\arcsec$) at which the flattened
     molecular core is detected.}
    \label{fig:pv}
  \end{center}
\end{figure}

\begin{figure}
  \scalebox{1.2}{\plotone{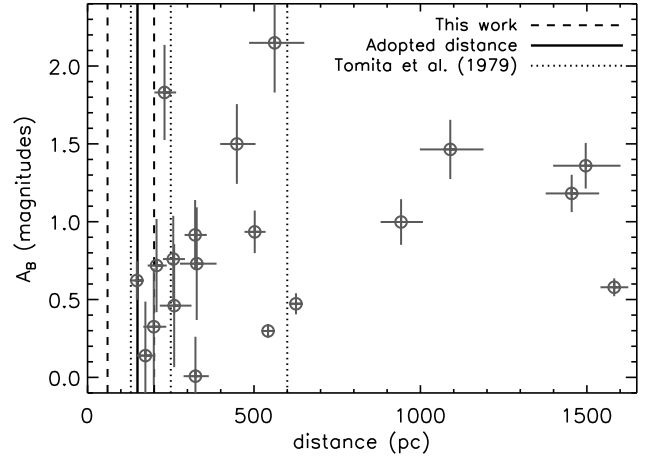}}
  \caption{$A_B$ vs.\ distance for stars within 20$\arcmin$ of \cbo.
    The error bars indicate the magnitude of the effect of a 15\%
    change in each direction in the $(B - K)$ color.  The dotted lines
    indicate the range in distances derived by \citet{tomita79}: 130,
    250 (adopted by those authors), and 600~pc, respectively.  Note
    that we do not reproduce a discontinuity in $A_B$ at 250~pc.  The
    dashed lines indicate plausible ranges in the distance based on
    the measured extent of the local bubble and the $A_B$ data shown
    here (see \S~3): 60 and 200~pc.  Our adopted distance of 150~pc is
    indicated by the solid line, and represents a conservative
    estimate.}
  \label{fig:dist}
\end{figure}

The $^{12}$CO and $^{13}$CO maps are presented in
Figures~\ref{fig:12co} through \ref{fig:pv}.  The large maps
($16\arcmin \times 12\arcmin$) show the full spatial extent
($\sim\!8\arcmin$) of the outflow as well as the velocity extent.  The
western side of the outflow is redshifted with respect to the line of
sight while the eastern side is blueshifted.  These maps are also used
by Shirley et al. (2008, in preparation) to calculate the outflow
opening angle.  They use this angle ($\sim\!55\degr$) to exclude from
their analysis of the submillimeter dust opacities regions where dust
may be affected by processes such as shock heating.  The $^{12}$CO and
$^{13}$CO lines are well resolved with 250~KHz
($\sim\!0.3$~km~s$^{-1}$) channel width; in $^{12}$CO the blueshifted
lobe is dominated by velocities ranging from $\sim\!5.04$ to
$\sim\!6.99$~km~s$^{-1}$ while the redshifted velocities range from
$\sim\!9.27$ to $\sim\!12.52$~km~s$^{-1}$ (see Figure~\ref{fig:12co}).
The subfield map (Fig.~\ref{fig:12cosub}) shows that the velocity
extent of the red- and blueshifted emission is in fact larger;
however, these maps do not cover the full $8\arcmin$ region of the
outflow.  The $^{13}$CO data do not show an outflow; the range of
these velocities is much narrower than the $^{12}$CO velocities, from
about 9.33 to 7.62~km~s$^{-1}$.  The $^{13}$CO data are dominated by
colder emission: in the large map (Fig.~\ref{fig:13co}) we see little
evidence of the outflow lobes observed in $^{12}$CO.  However, some
east-west elongation can be observed in the sub-field map, see
Figure~\ref{fig:13cosub}, due to the lower noise (rms
$\sim\!0.084$~K-T$_A ^*$) in the smaller map as compared to the full
field map (rms $\sim\!0.21$~K-T$_A ^*$).  Both large maps show an
extension of emission to the north and east of the protostar, on a
scale of $\sim\!9\arcmin$ in $^{12}$CO and $\sim\!6\arcmin$ in
$^{13}$CO, which is also observed in the 160~\micron\ mosaic.  The
outflow parameters calculated from the large area maps are presented
in Table~\ref{tab:co}.

The IRAC images reveal a resolved emission structure consisting of two
outflow lobes oriented in the east-west direction, seen clearly in
Figure~\ref{fig:mips} in all four IRAC channels.  The observation that
the eastern lobe is brighter than its western counterpart is explained
by the fact that the outflow is oriented slightly out of the plane of
sky; this orientation is corroborated by our CO maps in which the
eastern lobe is slightly blueshifted relative to the western lobe, as
discussed above.  It is also in agreement with other measurements: for
example, \citet{cabrit92} measured the inclination angle to the line
of sight to be $82\degr$.  The outflow not being aligned exactly on
the plane of the sky will give rise to differential extinction between
the eastern and western lobes, causing the western lobe to be more
obscured by the surrounding cold globule material out of which the
protostar was formed.

\begin{figure}[t]
  \scalebox{1.17}{\plotone{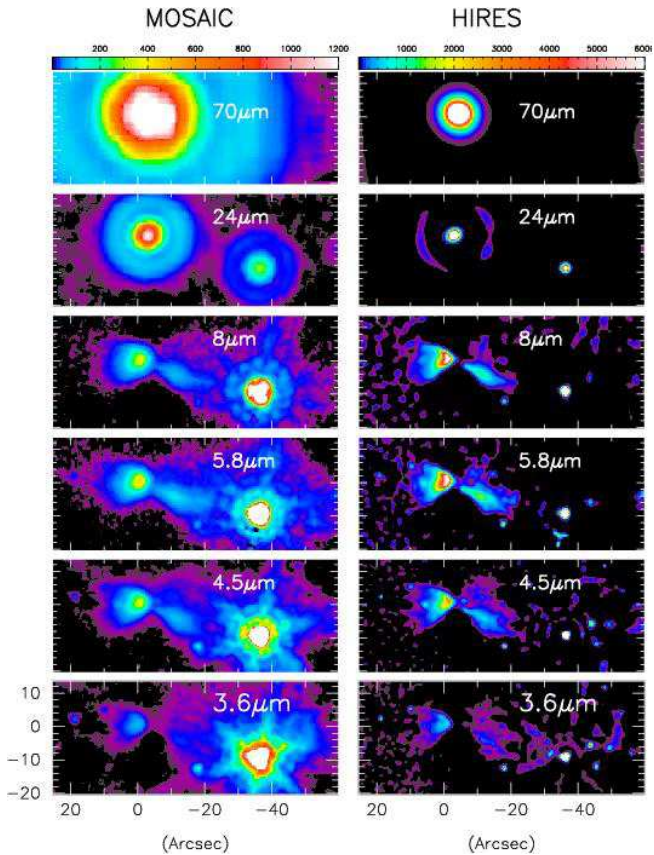}}
  \caption{Gallery of images of B335 (CB199) at the indicated
    wavelengths.  The images in the right column have been processed
    with HiRes for improved spatial resolution.  North is up and east
    is to the left.}
  \label{fig:hires}
\end{figure}

To improve the spatial resolution of the {\it Spitzer} data, we
present the deconvolved HiRes images of \cbo\ in
Figure~\ref{fig:hires} (right-hand column).  We reprocessed the raw
(BCD) data from the Spitzer archive, using a version of HiRes
deconvolution developed for Spitzer images by \citet{backus05} based
on the Richardson-Lucy algorithm \citep{richardson72,lucy74} and the
Maximum Correlation Method \citep{aumann90}.  HiRes deconvolution
improves the visualization of spatial morphology by enhancing
resolution (to sub-arcsec levels in the IRAC bands) and removing the
contaminating sidelobes from bright sources \citep{velusamy08}.  It
achieves a factor of 3 enhancement over the diffraction limited
spatial resolution from $\sim$ 1.22$\lambda$/D to $\sim$ 0.4
$\lambda$/D (e.g. from 2.4\arcsec to $\sim$ 0.8\arcsec at 8$\mu$m).
Typically the diffraction residue in the HiRes images is at levels
well below 0.1\% to 1\% for the IRAC and MIPS 24~\micron\ bands. The
Point-source Response Functions (PRFs), required as inputs to HiRes,
were obtained from the SSC; we used the extended (about 128 detector
pixel size in each direction) versions, supplied by Tom Jarrett
(private communication, 2007).  The observed images, in the form of
raw BCDs, were first post processed to fix contaminated data for
saturation, column pulldown, muxbleed, muxstriping, jailbars, and
outliers.  In the Richardson-Lucy algorithm the best resolution
enhancement is attained when the background is zero.  Accordingly, we
used the background subtracted BCDs.  In HiRes we used 50 iterations
for the IRAC bands and 100 for MIPS, which resulted in a FWHM of
0.55\arcsec to 0.75\arcsec for IRAC channels 1-4 and 1.6\arcsec at
24~\micron\ for a point source in the deconvolved images.  All bands
are relatively clean from artifacts except IRAC channel 4 (at 8$\mu$m)
which has some uncorrected muxbleed.  In the HiRes images diffraction
residues near the Airy rings are low (at a level $<$ 0.05\% of the
respective peak intensities at 3.6, 4.5, and 5.8~$\mu$m) providing
enhanced visualization closer to the protostar.

\begin{figure}[b]
  \begin{center}
    \scalebox{0.46}{{\includegraphics{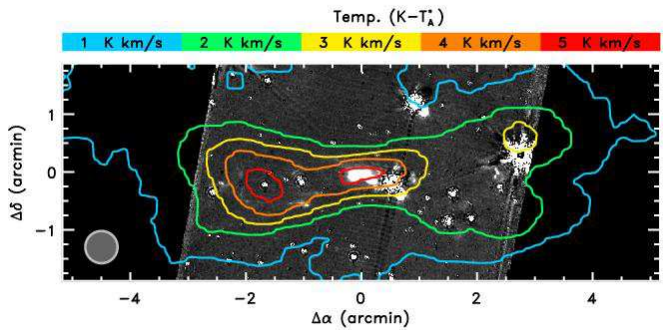}}}
    \caption{IRAC 4.5~\micron\ minus 3.5~\micron\ image with $^{12}$CO
      integrated contours at the indicated temperatures.  The
      countours are integrated over 0 to 15~km~s$^{-1}$.  The beam
      size ($32\arcsec$) is indicated by the grey circle.  Note the
      presence of HH objects, knots of dense shocked material
      producing molecular-H emission, located near the southern and
      eastern regions of the 4 K km/s contour.  North is up and east
      is to the left.}
    \label{fig:iracco}
  \end{center}
\end{figure}

In the HiRes 5.8~\micron\ image we measure the outflow opening angle
to be about $70\degr$, while in the deconvolved 8.0~\micron\ image we
measure the angle to be between $62\degr$ and $67\degr$.  Shirley et
al. 2008 (in preparation) measure the $^{12}$CO outflow opening angle
to be closer to $55\degr$.  In Figure~\ref{fig:iracco} we show the
$^{12}$CO integrated contours overlaid on the IRAC 4.5~\micron\ minus
3.6~\micron\ image.  It appears that the molecular outflow is more
confined than if it were simply expanding from the opening angle
inferred from the IRAC images.  Also note the presence of Herbig Haro
(HH) objects in the subtracted image along the inner walls of the
confining cavity, where it is probable that dense shocked gas produces
molecular hydrogen and CO emission that accounts for much of the
signal in the IRAC images.  These emission features fall in IRAC
channel 2.  The HH objects (some having a striking filamentary
geometry) imply that the inner cavity wall is quite sharp.  More
detailed discussion of the \cbo\ HH objects can be found in, e.g.,
\citet{galfalk07} and \citet{galvan04}.

The IRS spectrum (see Figure~\ref{fig:irs}) shows emission lines at
$\sim\!26$~\micron\ and $\sim\!35$~\micron\ which we identify as [FeII]
and [SiII], respectively.  These lines may be produced in the
interaction between the outflow and the surrounding neutral material.
We note that they are also seen in the spectrum of the north-east lobe
(HH47A) in the HH46/47 outflow, as discussed by \citet{noriega04} and
\citet{velusamy07}.  We also observe other features of interest in the
IRS spectrum, indicated in Figure~\ref{fig:irs}, like H$_2$O and
CO$_2$ ices and a deep silicate absorption feature
\citep{vandishoeck04}.

\subsection{The Shadow}

\begin{figure*}[t]
  \scalebox{0.8}{\includegraphics{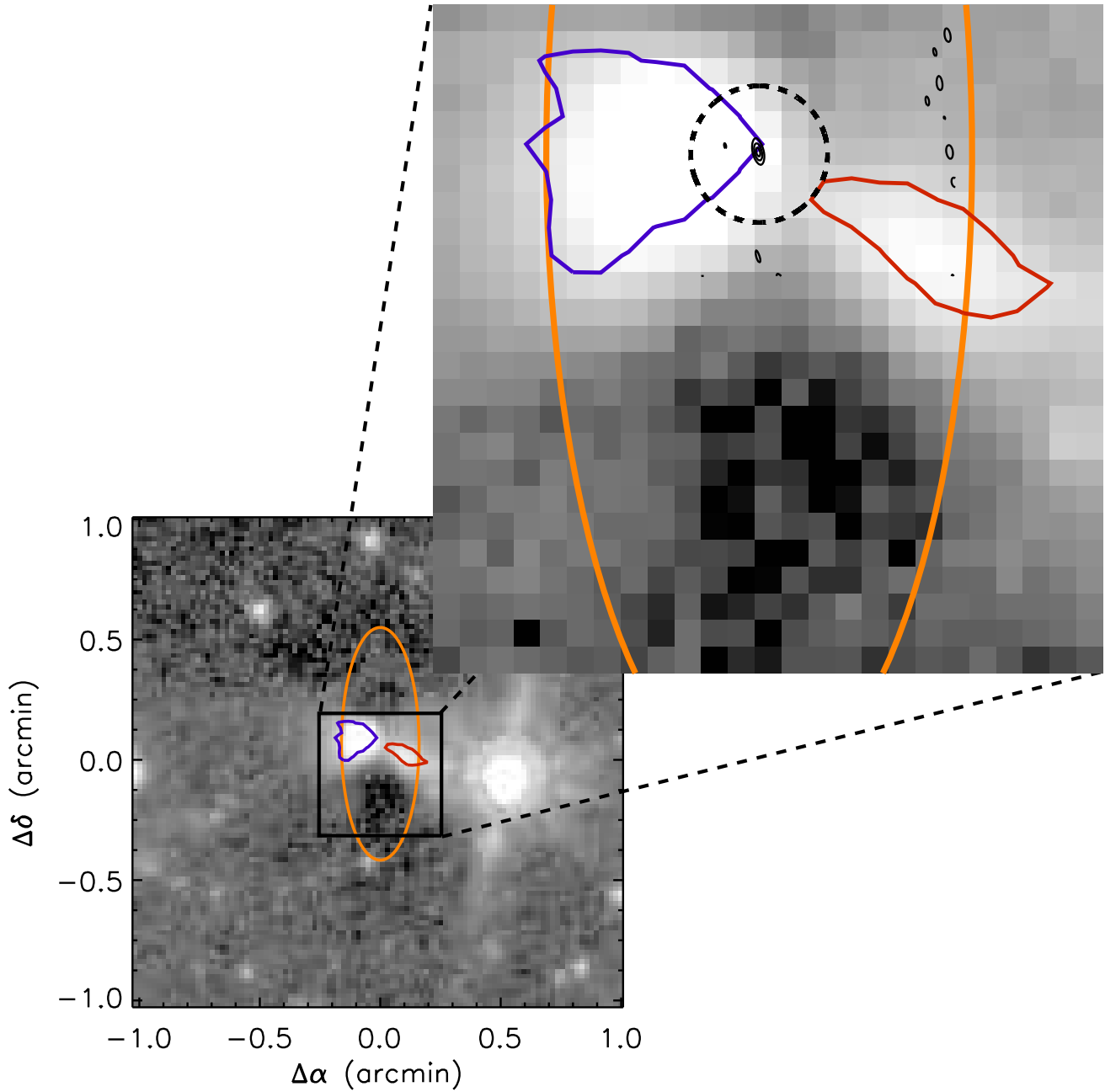}}
  \caption{IRAC 8.0~\micron\ image of \cbo, with inset, both shown on a
    log scale and at the original mosaic pixel scale of
    $1\farcs2$~pix$^{-1}$.  The 8.0~\micron\ shadow can be observed near
    the center and just south of the protostar.  We also show the
    HiRes 8.0~\micron\ deconvolved contours in red (west) and blue
    (east); these colors correspond to the CO outflow kinematics.  The
    orange ellipse has a 3:1 major to minor aspect ratio and indicates
    the rough size and shape of the flattened molecular core.  {\it
    Inset:} The solid black contours indicate the location of the
    circumstellar disk ($\sim$ 200~AU) observed at 1.2~mm by
    \citet{harvey03a}; the black dashed circle (with 3$\arcsec$
    radius) indicates the 24~\micron\ PSF.  North is up and east is to
    the left. }
      \label{fig:ch41mm}
\end{figure*}

The IRAC images show a shadow, located near the outflow lobes.  In
Figure~\ref{fig:mips} the shadow is evident at 3.6~\micron\ and
8.0~\micron, just to the south of the outflow lobes, as a dark nearly
circular feature about $20\arcsec$ in extent.  We also marginally
detect a northern counterpart in the 3.6~\micron\ and 8.0~\micron\ 
images (see Figures~\ref{fig:mips} and \ref{fig:ch41mm}).  We show the
IRAC images on a log scale, with the minimum and maximum image values
set to highlight the shadow and the outflow lobes; no other special
processing of the images was used to display the images.

\begin{figure}[t]
  \scalebox{1.15}{\plotone{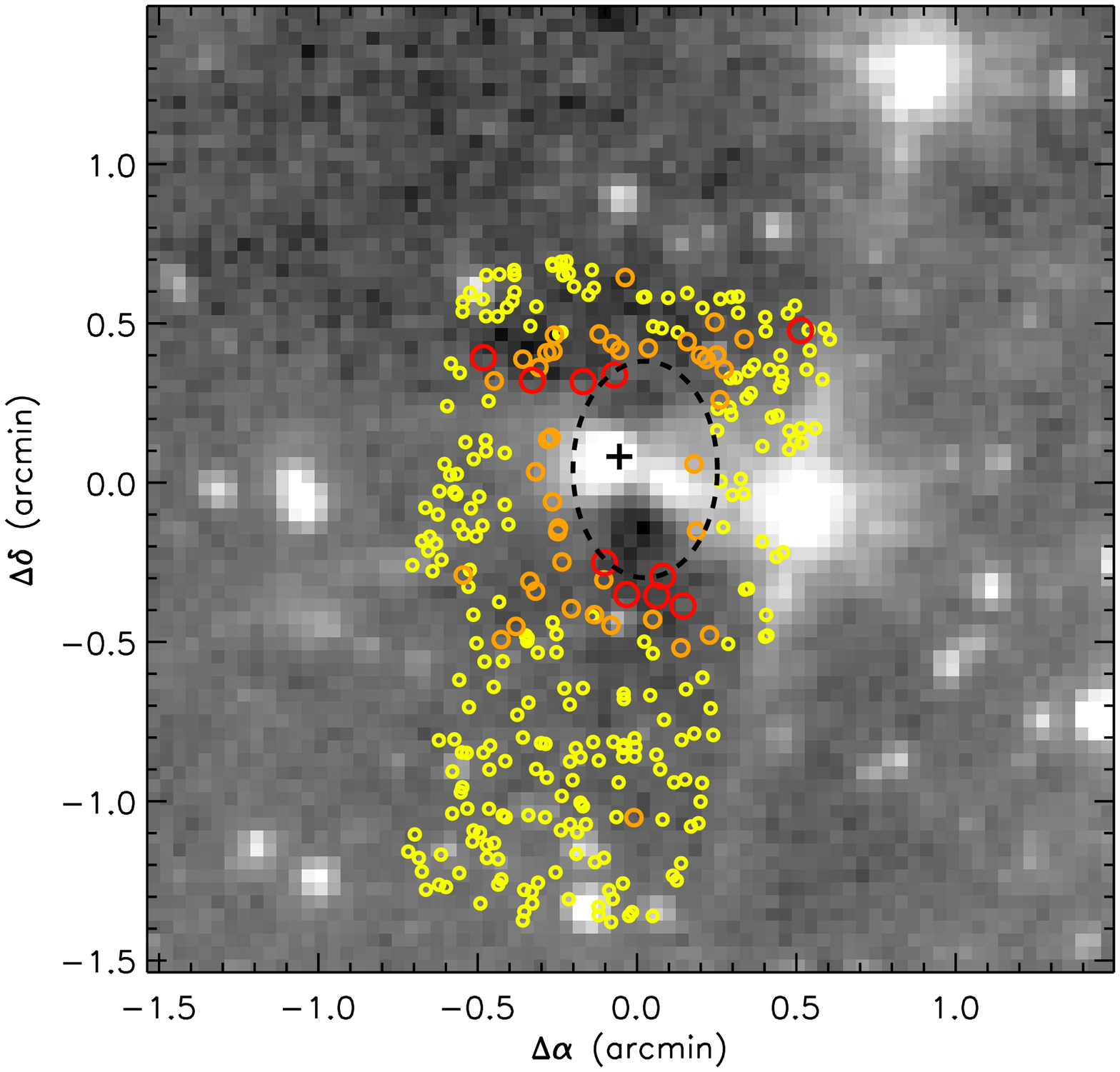}}
  \caption{IRAC 8.0~\micron\ image of \cbo, binned 2$\times$2 for a
    $2\farcs4$~pix$^{-1}$ resolution.  The 8.0~\micron\ shadow can be
    observed near the center and just south of the outflow lobes; a
    northern shadow counterpart is marginally dectected.  The circles
    indicate the locations of \citet{harvey01} near-IR stellar
    sources: the large red circles indicate sources with $(H - K)$
    color excesses greater than 2.9 (2$\sigma$, where for this set of
    observations $\sigma = 0.8$), the medium orange circles indicate
    sources with color excesses between 2.1 (1$\sigma$) and 2.9, and
    the small yellow circles indicate sources with excesses below
    2.1. The cross indicates the location of the \citet{harvey03a}
    detection of the central source. The dashed ellipse (shown with a
    3:2 axis ratio) is meant to highlight the shape of the inner
    region without stellar detections.  North is up and east is to the
    left.}
  \label{fig:ch4bin}
\end{figure}

The likely cause of the IRAC shadow is dense globule (dust and
molecular) material near the protostar.  The observed difference in
brightness between the east and west outflow lobes (apparent at all
IRAC wavelengths, see Figures~\ref{fig:mips} and \ref{fig:hires}), is
explained by: (a) the outflow orientation with the eastern lobe
pointing slightly towards the observer, and (b) the presence of dense
material at large radii (compared to the $\sim\!200$~AU circumstellar
disk) causing differential extinction between the two lobes.  This
interpretation is supported by the spatial distribution and measured
color excesses observed by \citet{harvey01}.  These data are presented
in Figure~\ref{fig:ch4bin}, where we show the locations of the stellar
sources (circles) overlaid on the binned 8.0~\micron\ image.  Here the
colors indicate a rough extinction distribution for the
\citet{harvey01} sources: red circles indicate sources with $(H - K)$
color excess greater that 2.9 (equal to the mean value plus $2\times$
the rms in the color excess distribution), the orange circles indicate
color excesses between 2.1 (equal to the mean value plus $1\times$ the
rms in the color excess distribution) and 2.9, and the yellow circles
indicate sources with excesses less than 2.1.  In this figure, the
ellipse (with a 3:2 major to minor axis ratio) indicates the rough
shape of the hole devoid of stars near the center of the cloud.  The
location of the most obscured sources around the outer circumference
of the southern shadow is suggestive that the shadow is in fact
tracing a flattened overdensity in molecular material.  On the
northern side the shadow is also detected in the IRAC image; its
location, like the southern counterpart, is bounded by background
stellar detections, indicating that the molecular core is in fact
flattened and that it has a significant degree of symmetry.

\begin{figure}[t]
  \scalebox{1.15}{\plotone{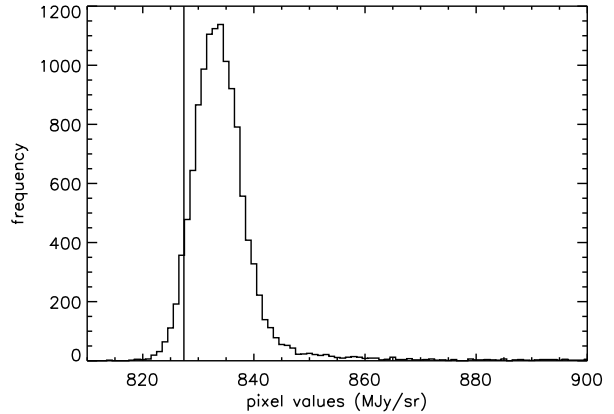}}
  \caption{IRAC 8.0~\micron\ pixel value distribution for a region
  $2\farcm2$ on a side, $2\farcm2$ north of B335.  The solid line
  indicates the 6th percentile value that we use for the image
  background value, equal to 827.408~MJy~sr$^{-1}$, chosen to get the
  best match to the stellar extinction values from \citet{harvey01}.}
  \label{fig:bg}
\end{figure}

\begin{figure}
 \scalebox{1.15}{\plotone{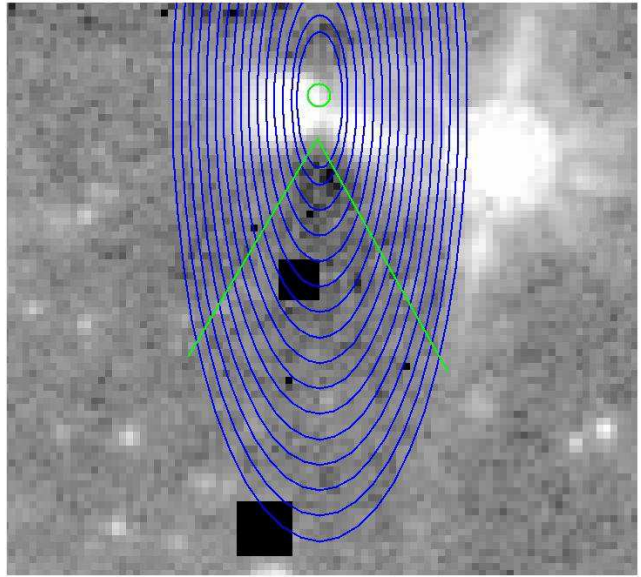}}
  \caption{IRAC 8.0~\micron\ image of B335, with two stars masked out
    (black boxes).  The blue ellipses, centered on the Harvey et
    al. location for the protostar (marked with a green circle)
    indicate the nested and adjacent regions we define to analyze the
    shadow.  The green lines indicate a $60\degr$ wedge between
    position angles 240$\degr$ and 300$\degr$, outside of which we
    reject all pixels due to light contamination from (a) the
    protostar reflection lobes, and (b) the diffraction spikes from
    the nearby very bright source.}
  \label{fig:reg}
\end{figure}

\begin{figure}
  \begin{center}
    \scalebox{0.55}{{\includegraphics{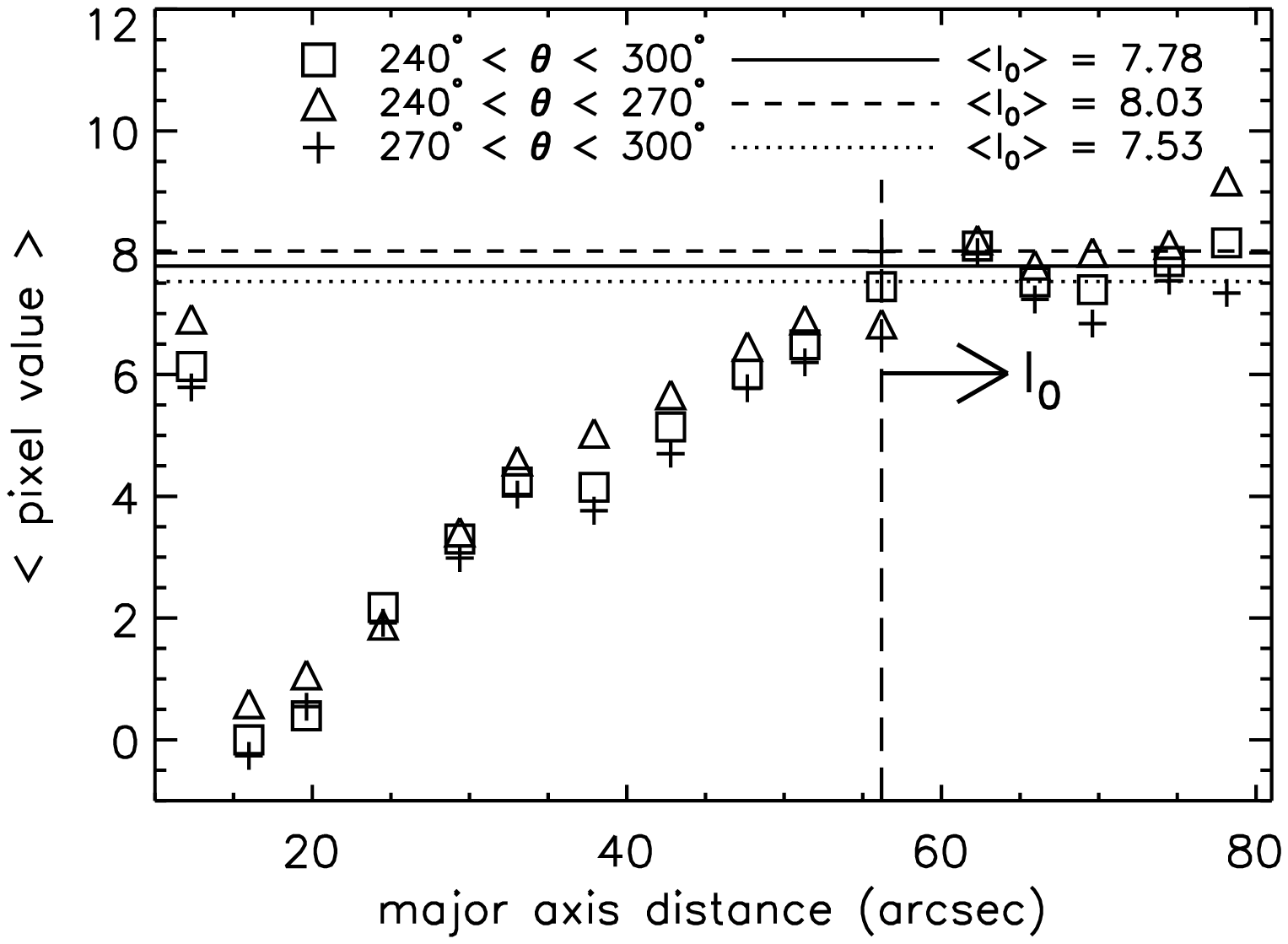}}}
    \scalebox{0.55}{{\includegraphics{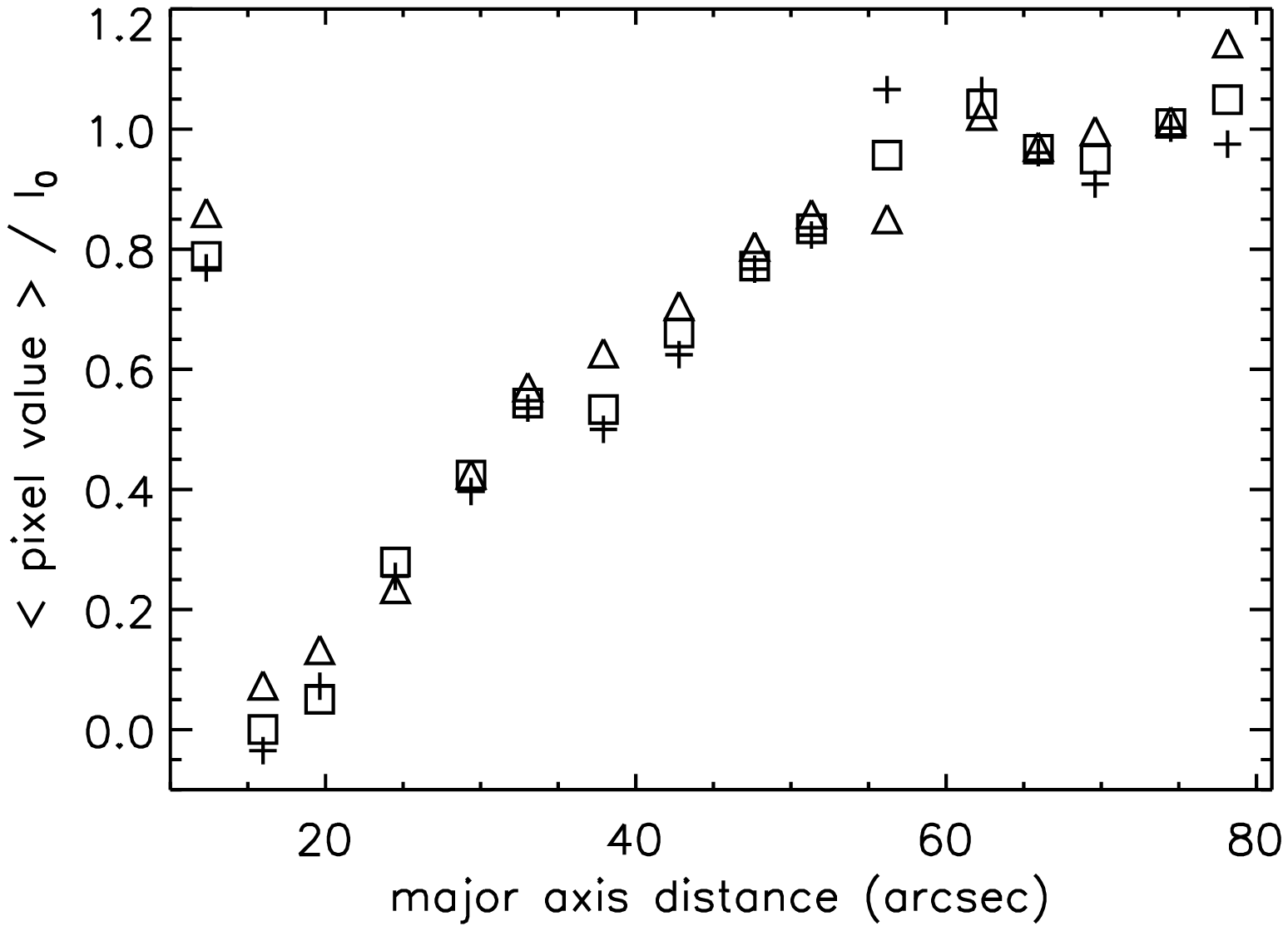}}}
    \caption{\small{IRAC 8.0\micron\ shadow profile.  {\it Top panel:}
      Average pixel values as a function of major axis distance from
      the \citet{harvey03a} location of the protostar.  Solid,
      short-dashed, and dotted lines indicate, respectively, the
      derived unobscured flux levels ($I_0$) in the three indicated
      regions: the full 60$\degr$ wedge, the eastern 30$\degr$ half,
      and the western 30$\degr$ half (located at the position angles
      indicated in teh top panel).  The vertical long-dashed line
      indicates the six outer regions used to derive $I_0$.  {\it
      Bottom panel:} $\exp(-\tau) = I/I_0$ as a function of major axis
      distance from the \citet{harvey03a} location of the protostar.
      {\it Both panels:} Squares indicate the profile derived using
      the full 60$\degr$ wedge, triangles indicate the eastern
      30$\degr$ half of the wedge, and crosses indicate the western
      30$\degr$ half of the wedge.  The bounding position angles of
      these wedges (240$\degr$, 270$\degr$, and 300$\degr$) are
      indicated in the top panel.  Note that there are not significant
      differences between the profiles, indicating that the derived
      profile is not strongly affected by the nearby bright source.}}
    \label{fig:pixval}
  \end{center}
\end{figure}

\begin{figure}[t]
  \scalebox{1.2}{\plotone{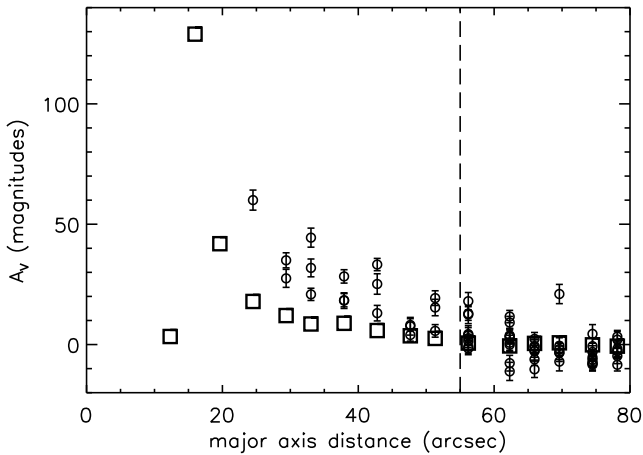}}
  \caption{$A_V$ values for the 8.0~\micron\ shadow (squares) compared
    to the \citet{harvey01} stellar $A_V$ values (circles) for each
    elliptical region.  We use the six outer radii --- indicated by
    the dashed line --- to normalize both the shadow and stellar
    extinction profiles.}
  \label{fig:shadow}
\end{figure}

In the text that follows we analyze the shadow and derive optical
depth and density profiles; we use techniques similar to those
presented in \citet{stutz07}.  The shadow is roughly centered on RA =
$19^h 37^m 01^s$, Dec = $+07^o 33\arcmin 55\arcsec$.  Although the
shadow is observed at both 3.6~\micron\ and 8.0~\micron\ we apply the
following analysis {\it only} to the 8.0~\micron\ shadow because the
3.6~\micron\ image has too many complicating factors, namely
contamination from the nearby bright source just west of \cbo\ and
many more stellar sources in and around the region of interest.  In
summary, to derive an 8.0~\micron\ profile we go through the following
steps: (1) re-normalize the image to our best estimate of the true
background value (Figure~\ref{fig:bg}), (2) divide the image up into
regions inside and outside of the shadow (Figure~\ref{fig:reg}), (3)
use these regions to derive optical depth, density, and extinction
profiles, all of which are equivalent, the latter two being subject to
an assumption about the dust properties at the wavelength of interest
(Figure~\ref{fig:pixval}), and (4), compare the derived profile to
available stellar color excesses, in this case the \citet{harvey01}
data, see Figure~\ref{fig:shadow}.  Here we discuss this process in
more detail.

The background level in the images, which we term $f_{DC}$ and which
is composed of instrumental background and zodiacal background, must
be measured and subtracted from the images to analyze the shadow.  We
estimate $f_{DC}$ by adjusting the level so that the depth of the
shadow and the extinction inferred from stellar observations agree.
We compare this level with the pixel values in a $2\farcm2$ squared
box located $2\farcm2$ north of the source selected for minimum
surface brightness; the distribution of pixel values is shown in
Figure~\ref{fig:bg}.  The chosen value for $f_{DC} =
827.40$~MJy~sr$^{-1}$ is the 6th percentile pixel value in the box.
The reasonably good agreement between our detected background and the
assumed one validates our procedure.

We divide the image into nested and concentric elliptical regions
shown in Figure~\ref{fig:reg}.  The elliptical regions are centered on
the \citet{harvey03a} location of the protostar, and have a
major-to-minor axis ratio of 3:1.  The ratio was chosen to match
roughly the location and shape of the 8.0~\micron\ shadow and to
satisfy the mininum requirements of the observed geometry: inspection
of Figure~\ref{fig:reg} provides the constraint that the flattened
core must lie in front of the entire western outflow lobe while still
providing the obscuration observed as a shadow at 8.0~\micron.  Smaller
axis ratios (rounder ellipses) will violate these constraints.
Similarly, larger axis ratios may allow for some of the emission from
the outer region of the western lobe to appear unobscured.  We reject
all pixels outside of a 60$\degr$ wide wedge pointed directly south;
this wedge is indicated as two green lines in Figure~\ref{fig:reg}.
It was chosen to avoid the diffraction spikes of the nearby bright
star and to mask out regions affected by emission from the outflow
lobes.  As can be seen in Figure~\ref{fig:reg}, we also mask out two
bright sources (black boxes) that would have affected the shadow
profile.  We then tabulate the mean pixel value in each elliptical
region, plotted in Figure~\ref{fig:pixval} (upper panel, square
symbols).  We use this distribution to calculate an estimate of the
optical depth, $\tau = -\ln (I/I_0)$, in the shadow.  To do so we must
measure $I_0$, the unobscured flux level near the shadow.  We measure
$I_0$ from the regions where the mean pixel values approach a
constant, in this case at elliptical major axis values greater than
56$\arcsec$.  We average over all pixel values outside this region and
inside 78$\arcsec$; we measure the value of $I_0 = 8.26$, indicated in
the upper panel of Figure~\ref{fig:pixval} as a solid line.  In the
lower panel of Figure~\ref{fig:pixval} we plot $\exp(-\tau_8) = I/I_0$
versus major axis distance.  One can clearly see that the inner-most
region is greatly affected by emission from the outflow lobes.

Motivated by the concern that emission from the bright source just to
the west of \cbo\ could be affecting our derived shadow profile we
perform the following test.  We divide the 60$\degr$ wedge between
position angles 240$\degr$ and 300$\degr$ into two 30$\degr$ wedges,
one on the east side, away from the bright source, and one on the west
side.  We proceed with the same analysis as described above and plot
the resulting profiles in Figure~\ref{fig:pixval}.  These profiles
(triangles indicate the east wedge, crosses indicate the west wedge)
do not show significant variations between each other, or relative to
the full wedge (again, indicated as squares).  Therefore the nearby
bright source is not significantly affecting our shadow profile.

Finally, using this $\tau$-profile, we derive an extinction profile.
Following the convention outlined in \citet{stutz07}, and assuming an
8.0~\micron\ dust opacity of $\kappa_{abs,8} = 8.13 \times
10^{2}$~cm$^2$~gm$^{-1}$ calculated by \citet{draine03a,draine03b}
from his $R_V = 5.5$ model, we derive the $A_V$ profile plotted in
Figure~\ref{fig:shadow}.  Here, the average mass column density in
each elliptical annulus is given by
\begin{equation}
\Sigma = \frac{\tau_8}{\kappa_{abs,8}}f,
\end{equation}
where $f (= 100)$ is the gas-to-dust ratio, and $\kappa_{abs,8}$ is
the absorption cross-section per mass of dust, and has the assumed
value noted above.  The choice of a model with a high $R_V$ value
relative to the diffuse ISM is intended to account for the grain
coagulation expected in dense, cold cores.  The extinction profile is
given by
\begin{equation}
A_{V} = \frac{\Sigma}{1.87\times 10^{21}\,{\rm cm}^{-2}\,{\rm
    mag}^{-1} \mu_{\rm H_2} m_{\rm H}}.
\end{equation}
We have assumed the conversion factor from the relation $N(H_2)/A_V =
1.87\times10^{21}$~atoms~cm$^{-2}$~mag$^{-1}$ \citep{bohlin78}, where
we assume that all of the hydrogen is in molecular form.  We have
adopted $R_V = 3.1$ --- a value appropriate for the diffuse ISM where
this relation was measured --- to convert the \citet{bohlin78}
relation from $E(B-V)$ to $A_V$.  At longer wavelengths than $V$ the
extinction law does not appear to be a strong function of density.
Here $\mu_{\rm H_2} = 2.8$ is the effective molecular weight per
hydrogen molecule.  For this set of constants $A_V = 14.04 \times
\tau_8$.  Following a different method, where the conversion is
obtained directly from the \citet{draine03a,draine03b} dust models,
$A_V = \kappa_{abs,V}/\kappa_{abs,8}\times A_8 = 9.7\times \tau_8$.
The two $A_V$ determinations are consistent within the uncertainties
of the measured infrared extinction values (see below).  To take
possible freezeout and grain growth effects into account we consider
the \citet{ossenkopf94} model opacities; for models generated at a
density of $n = 10^6$~cm$^{-3}$ these authors derive 8.0~\micron\
opacities of $\kappa_{abs,8} = 1.06 \times 10^{3}$~cm$^2$~gm$^{-1}$
and $\kappa_{abs,8} = 1.21 \times 10^{3}$~cm$^2$~gm$^{-1}$ for thin
and thick grains, respectively.  These opacities will diminish the
derived $A_V$ values by factors of 1.3 and 1.5 respectively.  We note
that the range in measured infrared extinction
\citep[e.g.,][]{indebetouw05,flaherty07} is significant; the $A_V$
determination is therefore uncertain by a factor $\sim\!2$ and as a
result so is the column density.

Figure~\ref{fig:shadow} shows the derived shadow extinction profile
(square symbols); we compare our profile to the values derived from
the \citet{harvey01} stellar data in each elliptical region.  To make
use of the stellar observed $(H-K)$ colors we follow the procedure in
\citet{harvey01}.  In addition to the statistical uncertainty in
$(H-K)$, they also measured the uncertainty due to the intrinsic
scatter in $(H-K)$ of background stars determined from a nearby
``OFF'' field.  The mean intrinsic $(H - K) = 0.13 \pm 0.16$ for the
``OFF'' field.  To determine the color excess, the stellar $(H-K)$
colors have 0.13 subtracted from them; to determine the total errors
we add $0.16$ in quadrature to the statistical errors.  Finally, we
use $A_V / E(H-K) = 15.87$ to convert to $A_V$ \citep{rieke85}.  To
make a fair comparison between the shadow profile and the stellar
color exess, we must subtract off the mean value of the stellar
extinction in the regions used to find $I_0$ for the shadow, i.e. we
account for the differences in the two zero-points for the profiles.
In this case, $\langle A_V \rangle = 14.52$, or $E(H-K) = 0.92$, for
the stars in this region.  

The maximum extinction level derived for the shadow is $A_V \sim 100$.
Our spectrum (Fig.~\ref{fig:irs}) shows silicate absorption with a
$\tau \gtrsim 2.5$.  Because of possible complexities with
interpreting this spectrum, we consider $\tau \sim 2.5$.  Applying a
standard conversion \citep{rieke85} the corresponding extinction is
$A_V \gtrsim 40$.  We interpret the measurement as a lower limit
because radiative transfer in a disk with a temperature gradient will
tend to fill in the intrinsic feature.  If we assume an equal amount
of material behind the star, the limit on the total column is
equivalent to $A_V \gtrsim 80$.  Thus, the spectrum is consistent with
our shadow analysis: there is a column density of N(H$_2$)$\sim
10^{23}$~cm$^{-2}$ or more towards the protostar.  For a radius of
$20\arcsec$, this column yields a mass $\sim\!0.5$~\msun.

Figure~\ref{fig:ch41mm} shows additional components of the source in
relation to the shadow.  \citet{harvey03a} used the IRAM
interferometer in the A configuration, which has baselines of 43 to
400~m.  At this high resolution they detect the protoplanetary disk in
emission, with a maximum diameter of about $1\arcsec$ (150~AU, see
Fig.~\ref{fig:ch41mm}), but resolve out larger structures.
\citet{harvey03b} used the D configuration with baselines of 15 to
80~m, which allow for larger structures to be resolved.  Their best
fit model to the data (p = 1.65, $\alpha $ = 40$\degr$, see their
Figure 3) shows a structure about $7\arcsec$ in extent aligned in the
north-south direction (not shown in Fig.~\ref{fig:ch41mm}).  This
emission likely originates from the innermost material associated with
the 8.0~\micron\ shadow.  The dashed circle indicates the 24~\micron\
PSF ($\sim\!3\arcsec$ in radius) which is coincident with the
protostar and disk.

Our analysis does not include any treatment of the possible foreground
emission that may arise from the surface of the globule its self; this
emission, if present, would have the net effect of causing an
underestimate of the derived optical depth.  A rigorous modeling of
this emission component is beyond the scope of this paper.  However,
inspection of Figure~\ref{fig:shadow} indicates that if this effect is
present it is not large; the \citet{harvey01} stellar data lie
preferentially above our shadow profile, at higher $A_V$'s.
Foreground emission could explain this systematic offset.  However,
other uncertainties, discussed above, are likely to dominate our
extinction profile estimate.

\subsection{Detection of a flattened molecular core}

The sub-field $^{12}$CO and $^{13}$CO channel maps show a north-south
structure that appears to be rotating, seen in
Figures~\ref{fig:12cosub} and \ref{fig:13cosub}.  This structure
extends out to a radius of about 80$\arcsec$ (12000~AU) from the
protostar.  In Figure~\ref{fig:12cosub} the southern component is
detected between roughly 8.94~km~s$^{-1}$ and 9.92~km~s$^{-1}$; at
8.94~km~s$^{-1}$ the image is dominated by the center velocity
emission.  The northern counterpart is weaker in emission, but is
apparent at 7.31~km~s$^{-1}$ and fades at velocities smaller than
6.66~km~s$^{-1}$.  The maximum relative velocity between the two
sides is therefore about 3~km~s$^{-1}$.  The $^{13}$CO map shows
little indication of this velocity gradient; however, there is a
north-south elongation of the emitting region.  Additionally, CO
freeze-out may complicate the analysis of the molecular data.  The
difference in appearance in $^{12}$CO between the northern and
southern regions can be explained by clumpy material in a rotating
structure, expected by models (see below).  Thus, both our extinction
shadow image and our CO line mapping indicate that the dense core of
the globule is flattened and may be rotating around the protostar.

There have been many indications of this flattened structure
previously, on different angular scales depending on the type of
observation. For example, the 800~\micron\ image by \citet{chandler90}
indicates a low surface brightness north-south extension of about
2$\arcmin$, compared with only about 1$\arcmin$ for the east-west
extension.  \citet{velusamy95} find a very similar structure in their
filled aperture image in CCS at 22 GHz.  They also present channel
maps that show no convincing evidence of overall rotation.
\citet{hodapp98} shows deep images at H and K that reveal shadows of
the cloud against the diffuse light from the background, a technique
analogous to ours at 8.0~\micron. At H, the shadow is slightly
flattened, but it is significantly more so at K-band. This change is
probably due to the ability of the K-band image to penetrate larger
optical depths than at H-band.  \citet{saito99} show a similar
structure in C$^{18}$O and find marginal evidence that it is rotating
with a small redshift to the north and blue shift to the south.  The
indicated rotation is only about 0.05 km/sec over the range of
2$\arcmin$ (0.09pc).

\citet{wilner00} observed in CS J = 5-4 with the IRAM interferometer
and an angular resolution of 2.5$\arcsec$; their channel maps do not
appear to show the full structure, but there is a signal near the
systemic velocity that is well resolved with a length of $5\arcsec$ to
$6\arcsec$ oriented approximately N-S. This structure is also seen in
the IRAM interferometer continuum data obtained at 1.2mm
simultaneously \citep{harvey03b} and by \citet{jorgensen07} with the
SMA (the latter image also shows a faint extension to the north and
another to the SSW). It is plausible that the latter three references
underestimate the extent of the larger, low surface bright structure
because it is partially resolved out with the interferometers.  This
possibility is partially supported by the strong detection of the
north-south structure over a length scale of $2\arcmin$ in the filled
aperture CS observation by \citet{velusamy95}.

Three-dimensional numerical simulations of cloud collapse are growing
rapidly in sophistication. In general, they show that the collapse
proceeds quickly into filaments, or disks with spiral arms, and with
typical dimensions of 1000 AU (diameter)
\citep{whitehouse06,krumholz07,arreaga07}, a result that was also
anticipated in some earlier work \citep[e.g.,][]{nelson97}.  In the
simulations, these structures often break up into binary or multiple
forming stellar systems with dense and more compact
circumstellar/protoplanetary disks. In any case, the interstellar
medium becomes organized into two distinct classes of structure: 1.)
an evolving, pseudo-stable protoplanetary disk of size up to a couple
of hundred AU; and 2.) what we term a {\it flattened molecular core}
to include the variety of filaments, disks, and spiral arms produced
by the simulations, and of typical size 1000 AU. Viewed edge-on, most
versions of flattened molecular cores would appear as linear
structures similar to edge-on disks. By an age of order 100,000~years
or less, the natal cloud dissipates to reveal the new protostar, and
the flattened molecular cores have disappeared.  The protostellar disk
remains, within a typical 200 AU diameter \citep[see ][]{andrews07}.
In turn, it evolves and dissipates on a time scale of a few million
years, leaving a planetary system with typical dimensions of 200 AU.

The \cbo\ structure would appear to be similar to the elongated shadow
in the IRAC images of L~1157 discovered by \citet{looney07}. However,
the L~1157 shadow is $\sim$ 15,000 to 30,000~AU in diameter, somewhat
larger than that around \cbo.  \citet{looney07} suggest that it
represents a flattened density enhancement produced as a result of the
general influence of magnetic fields or rotation in the collapse of a
cloud; that is, it is part of the general trend that yields a
flattened molecular core, but probably on a size scale that decouples
much of its volume from direct participation in the formation of the
central protostar.

\begin{deluxetable}{lccr}
\tabletypesize{\scriptsize}
\tablecaption{Mass Estimates}
\tablewidth{0pt}
\tablehead{
\colhead{Data} 
& \colhead{Temp.}
& \colhead{Radius}
& \colhead{Mass [$\msun$]}\\
\colhead{  }
& \colhead{[K]} 
& \colhead{[$\arcsec$]} 
& \colhead{[$\msun$]}\\
\colhead{(1)}
& \colhead{(2)} 
& \colhead{(3)} 
& \colhead{(4)} 
}
\startdata
8.0~$\micron$        &  \nodata   &     20   &     0.5  \\
160~$\micron$        &       10   &     48   &    43    \\
160~$\micron$        &       10   &     80   &    69    \\
160~$\micron$        &       12   &     48   &    10    \\
160~$\micron$        &       12   &     80   &    15    \\
160~$\micron$        &       14   &     48   &     3.2  \\
160~$\micron$        &       14   &     80   &     5.2  \\
$^{13}$CO            &       30   &    300   & $\geqslant$0.08$^a$ \\
$^{13}$CO            &       10   &    300   & $\geqslant$0.2$^a$ \\
$^{12}$CO$^b$        & \nodata    &      80  &     35   \\
\enddata
\label{tab1}
\tablenotetext{a}{Possible CO freezeout would imply that this value is
  a lower limit.}
\tablenotetext{b}{Enclosed mass, calculated assuming a circular velocity $=
  1.63$~km~s$^{-1}$ (see \S4.1)}
\end{deluxetable}

\subsection{Globule Mass}

A number of lines of evidence suggest that the full globule has
substantial mass.  We find $\sim$ 0.5 M$_\odot$ within a $40\arcsec$
beam, based on the absorbing column of material.  \citet{gee85}
modeled the spectral energy distribution (fitting it with a 14~K
temperature and a $\nu^2$ emissivity) to find a mass that translates
to $\sim$ 4~\msun\ at our preferred distance of 150~pc. Similar masses
were derived by \citet{keene83} and \citet{davidson87}, while somewhat
smaller values (by factors of $\sim$ 3) have been derived by
\citet{walker90} and \citet{shirley02}.  Correcting these measurements
to a scale comparable to the $^{12}$CO source diameter (e.g.,
according to the relative flux density versus aperture at 160$\mu$m)
increases them by a factor of 2 to 3.

At the same time, the calculations depend critically on the assumed
temperature of the dust.  An upper limit of $\sim$ 14K can be adopted
on the basis of the SED \citep{gee85}, plus expectations for dust
temperatures due to heating by the ISRF \citep{deluca93}.  In Table~3,
we show the dependence of calculated mass on dust temperature within
this constraint, both for a 40" aperture and for the full source.  We
conclude that the mass of the entire globule is likely to be greater
than 5~\msun, perhaps even by a factor of a few.

A total cloud mass of 10~\msun, if rotationally supported against
collapse, would have a net velocity gradient of
$\sim\!1.5$~km~s$^{-1}$.  Different molecular lines give differing
measures of the velocity gradient in the cloud, from zero to
$\sim\!3$~km~s$^{-1}$ (the latter from our $^{12}$CO data).  From
these discrepancies it is likely that the cloud is not simply in
smooth rotation but probably has a complex structure in terms of both
density and dynamics \citep[e.g.][]{zhou90}.  Nonetheless, the overall
gradient we observe in $^{12}$CO supports the argument for a mass of
$\sim\!10$~\msun\ or more.

\section{The Protostar}

Figure~\ref{fig:mips} shows our three MIPS \cbo\ mosaics.
Figure~\ref{fig:hires} shows the deconvolved HiRes 24 and 70~\micron\
images.  In Figure~\ref{fig:ch41mm} we indicate the location of the
protostar from \citet{harvey03a} with black contours (RA = $19^h 37^m
0.89^s$, Dec = $+07^o 34\arcmin 10.9\arcsec$).  The dashed circle
indicates the 24~\micron\ PSF size, which is well centered on the
radio continuum measurement.  The 70~\micron\ PSF is well centered on
the protostar, while the 160~\micron\ point-source is slightly offset
most likely due to artifacts in the mosaic.  The SCUBA 450 and
850~\micron\ observations of \cbo\ \citep{shirley00} are also well
centered on the radio observations.  This object thus coincides with a
bright point source seen in each of the MIPS bands.  As illustrated by
Figure~\ref{fig:ch41mm}, the protostar is invisible in IRAC channel 4
(and all other IRAC channels).  However, in the MIPS bands it
dominates the emission.  We detect a small asymmetry in the
24~\micron\ PSF in the east-west direction.  This asymmetry becomes
more evident in the HiRes deconvolved images.  The elongation is
clearly seen along the east-west direction after a PSF subtraction to
take out the bright symmetric core of the PSF.  We note that the
direction of the 24~\micron\ PSF asymetry (PA = $\sim\!100\degr$) is
somewhat misaligned with the IRAC emission (see Fig.~\ref{fig:hires})
but in agreement with the $^{12}$CO outflow orientation (see
Fig.~\ref{fig:iracco}).  The protostar accounts for almost all of the
70~\micron\ radiation and for the point source seen at 160~\micron\
as well.  As noted previously, the 160~\micron\ emission is also
extended to the north-west of the YSO.  This extension is in the
direction of the main body of globule, and can be seen in our
$^{12}$CO and $^{13}$CO maps (see Figures~\ref{fig:12co} and
\ref{fig:13co}).

By combining the measurements across the Spitzer bands with those at
longer wavelengths from the Kuiper Airborne Observatory (KAO) and
ground-based telescopes, we have produced the composite spectral
energy distribution for B335 shown in Figure~\ref{fig:irs}.  The
Spitzer images show that the infrared emission from \cbo\ is confined
spatially out to 70~\micron.  At 160~\micron, the images show a
compact source plus a diffuse component; these results are consistent
with those of Keene (1983) showing that the flux from the source
starts increasing with measurement aperture only longward of
140~\micron.

In Table~\ref{tab:phot} we list the present photometric results as
well as selected results taken from the literature.  IRAC data are
tabulated at both $2\farcs4$ and $24\arcsec$ aperture diameters to
facilitate the model comparison discussed below.  At 160~\micron\ and
beyond we also give flux densities in two apertures.  The flux density
at 160~\micron\ is 68~Jy in a $50\arcsec$ beam and 156~Jy into a
$180\arcsec$ beam.  The flux measured at 70 and 160~\micron\ is
consistent within the uncertainties with the earlier results published
by \citet{keene83} in this wavelength range based on measurements from
the KAO.  The flux density measurements at 350~\micron\ and beyond are
measured at $40\arcsec$ beams; measurements with a $180\arcsec$ beam
are not available at these submillimeter wavelengths.  To obtain a
self-consistent estimate of the total flux radiated in a $180\arcsec$
beam, we scale the flux measured at 350~\micron\ and beyond into a
$40\arcsec$ aperture by a factor of 2.3, which is the factor by which
the 160~\micron\ flux increases between the two aperture sizes.  This
procedure does not contradict any of the larger beam measurements
beyond 350~\micron.  In addition, as the luminosity in a $\nu {\rm
F}_{\nu}$ sense is dominated by the 160~\micron\ luminosity, our
estimates will be valid.

\begin{figure}
  \scalebox{1.2}{\plotone{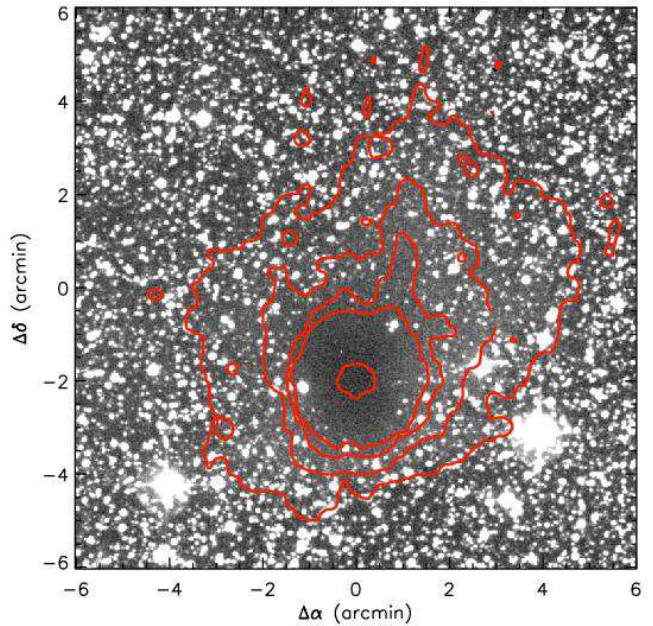}}
  \caption{Red Digital Sky Survey (DSS) image with 160~\micron\
  contours overlaid in red.  The DSS image is displayed on a linear
  scale, centered on RA = $19^h 37^m 00.5^s$, Dec = $+07^o 36\arcmin
  10.0\arcsec$ ($2\arcmin$ north of the protostar).  Contour levels are
  $\{0.2,0.5,0.8,1.1,10\}\times1000$~mJy/arcsec$^2$.}
  \label{fig:dss}
\end{figure}

Integrating over the SED plotted in Figure~\ref{fig:irs} for the point
source component, we find a luminosity of $3.2$~L$_\sun$ at an assumed
distance of 250~pc, consistent with previous results.  Adopting a
distance of 150~pc, the luminosity becomes 1.2~L$_\sun$.  The larger
beam data give a luminosity over a $180\arcsec$ region of
1.9~L$_\sun$, peaking at 160~\micron, for a distance of 150~pc.  The
difference, 0.7~L$_\sun$, is consistent with heating by the
interstellar radiation field: the interstellar radiation density is
given by \citet{cox89} as 0.5~eV~cm$^{-3}$.  The interstellar flux
incident on a sphere of radius $\sim\!0.07$ pc (corresponding to
$90\arcsec$ at a distance of 150~pc) is, coincidentally, 0.8~L$_\sun$.
The opacity of \cbo\ is such that we can expect most of this flux to
be absorbed and reradiated in the far infrared, thus producing diffuse
emission from the cloud at a level consistent with our observations
(see Fig.~\ref{fig:dss}).

Detailed modeling of the present data set is deferred to a subsequent
work.  However, to place \cbo\ into context, we have compared its
properties with those of the grid of models recently made available on
the web by \citet{robitaille06} and \citet{robitaille07}.  These
models for young protostars include disk and envelope components,
scattered light, and outflow driven cavities.  The SED is computed for
a range of viewing angles, and the distance to the source can be
varied as well.

We input into the SED modeler the stellar mass range of 0.3 to
0.5~\msun.  The range of selected models is plausible and the general
shape of the model SED's match the observed photometry.  We also fit
the observed photometry (shown in Table~2) with a range of distances
from 100~pc to 300~pc.  The fitter prefers models with lower distances
and high inclinations, which corroborates our distance estimate and
agrees with the observed geometry of the source.  The best-fit model
has a $\chi^2 = 561$, a mass of 0.3~\msun, a distance of 100~pc, an
inclination $\sim\!87\degr$, and a disk outer radius of 145~AU, in
good agreement with the \citet{harvey03a} measurement.  If \cbo\ is
placed at the traditional distance of 250~pc, the calculated
luminosity exceeds that expected from a $\sim\!0.5$~M$_\odot$
protostar and it is difficult to find a consistent solution without
invoking asymmetric output of the far infrared flux. The set of 10
models with lowest $\chi^2$ values generally agree with this best-fit
scenario, with stellar masses ranging from $\sim\!0.2$ to 0.7~\msun.
Masses derrived from fitting SEDs to models are currently very
uncertain; however, these masses fall in an interesting range and are
only $\sim\!5$\% of the total globule mass.

\section{Conclusions}

We report and analyze new Spitzer 3.6 to 160~\micron\ data
observations of \cbo\, along with new ground based data in $^{12}$CO
and $^{13}$CO.  

\noindent 1.)  We revise the estimated distance from 250~pc to
150~pc.  Our preferred distance is more compatible with the extinction
models of \citet{harvey01} and with models of the embedded
protostar \citep{robitaille06,robitaille07}.

\noindent 2.)  We discover an 8.0~\micron\ shadow and derive a density
profile for radii ranging from $\sim\!3,000$ to 7,500 AU. The shadow
opacity at maximum is equivalent to $A_V$ = 100 or more.

\noindent 3.)  We detect a rotating structure in $^{12}$CO on scales
of $\sim\!10000$~AU which we term a flattened molecular core.  This
structure is coaxial with the central circumstellar disk.

\noindent 4.)  It is very likely that the 8.0\micron\ shadow and the
flattened molecular core originate from the same structure.

\noindent 5.)  We measure the luminosity of the protostar to be
1.2~L$_\odot$ (at 150~pc),in good agreement with models for young
stars of the expected mass ($\sim\!0.4$~\msun) and general
characteristics of the object.

As the prototypical isolated, star-forming dark globule, there is a
huge literature on Barnard 335.  Nonetheless, our mid- and
far-infrared data from {\it Spitzer} and CO observations from the HHT
provide new insights regarding the configuration of the collapse of
the globule and the properties of the protostar. We have found that
the new observations are roughly consistent with recent models.
Further comparisons with the full suite of observations have the
potential to improve our understanding of isolated star formation
significantly.

\acknowledgments We thank Kevin M. Flaherty, Fabian Heitsch and Craig
Kulesa for helpful comments.  We thank Thomas Robitaille for his help
with model fitting.  The authors also thank the anonymous referee for
helpfull comments which improved the text.  A.M.S. thanks the c2d team
members for insightful comments.  Portions of this work were carried
out at the Jet Propulsion Laboratory, California Institute of
Technology, under contract with the National Aeronautics ans Space
Administration.  This work was supported by contract 1255094 issued by
Caltech/JPL to the University of Arizona.  This work was supported in
part by National Science Foundation grant AST-0708131 to The
University of Arizona.  MK was supported by the KRF-2007-612-C00050
grant.


\begin{thebibliography}{}

\bibitem[Andrews \& Williams(2007)]{andrews07} Andrews, S.~M., \&
Williams, J.~P.\ 2007, \apj, 659, 705

\bibitem[Arreaga-Garc{\'{\i}}a et al.(2007)]{arreaga07}
Arreaga-Garc{\'{\i}}a, G., Klapp, J., Sigalotti, L.~D.~G., \&
Gabbasov, R.\ 2007, \apj, 666, 290

\bibitem[Aumann et al.(1990)]{aumann90} Aumann, H.~H., Fowler, 
J.~W., \& Melnyk, M.\ 1990, \aj, 99, 1674 

\bibitem[Backus et al.(2005)]{backus05} Backus, C., Velusamy, T.,
Thompson, T., \& Arballo, J.\ 2005, Astronomical Data Analysis
Software and Systems XIV, 347, 61

\bibitem[Bertin \& Arnouts(1996)]{bertin96} Bertin, E., \& Arnouts, S.\
1996, \aaps, 117, 393

\bibitem[Blanco et al.(1970)]{blanco70} Blanco, M., Demers, S., \&
Douglass, G.~G.\ 1970, Publications of the U.S.Naval
Observatory.~Second Series, Washington: United States Government
Printing Office (USGPO), 1970

\bibitem[Bohlin et al.(1978)]{bohlin78} Bohlin, R.~C., Savage, B.~D.,
\& Drake, J.~F.\ 1978, \apj, 224, 132

\bibitem[Bok \& Reilly(1947)]{bok47} Bok, B.~J., \& Reilly, E.~F.\
1947, \apj, 105, 255

\bibitem[Bok(1948)]{bok48} Bok, B.~J.\ 1948, Centennial Symposia, 53

\bibitem[Bouigue et al.(1961)]{bouigue61} Bouigue, R., Boulon, J., \&
Pedoussaut, A.\ 1961, Annales de l'Observatoire Astron.~et Meteo.~de
Toulouse, 28, 33

\bibitem[Cabrit \& Bertout(1992)]{cabrit92} Cabrit, S., \& Bertout,
C.\ 1992, \aap, 261, 274

\bibitem[Chandler et al.(1990)]{chandler90} Chandler, C.~J., Gear,
W.~K., Sandell, G., Hayashi, S., Duncan, W.~D., Griffin, M.~J., \&
Hazella, S.\ 1990, \mnras, 243, 330

\bibitem[Choi et al.(1995)]{choi95} Choi, M., Evans, N.~J., II,
Gregersen, E.~M., \& Wang, Y.\ 1995, \apj, 448, 742

\bibitem[Choi(2007)]{choi07} Choi, M.\ 2007, \pasj, 59, L41 

\bibitem[Clemens \& Barvainis(1988)]{clemens88} Clemens,
D.~P., \& Barvainis, R.\ 1988, \apjs, 68, 257

\bibitem[Cox \& Mezger(1989)]{cox89} Cox, P., \& Mezger, P.~G.\ 1989,
\aapr, 1, 49

\bibitem[Davidson(1987)]{davidson87} Davidson, J.~A.\ 1987,
\apj, 315, 602

\bibitem[de Luca et al.(1993)]{deluca93} de Luca, M., Blanco, A., \&
Orofino, V.\ 1993, \mnras, 262, 805

\bibitem[Dickman(1978)]{dickman78} Dickman, R.~L.\ 1978, \apjs, 37,
407

\bibitem[Draine(2003a)]{draine03a} Draine, B.~T.\ 2003a, \araa, 41,
241

\bibitem[Draine(2003b)]{draine03b} Draine, B.~T.\ 2003b, \apj, 598,
1017

\bibitem[Evans et al.(2005)]{evans05} Evans, N.~J., II, Lee, 
J.-E., Rawlings, J.~M.~C., \& Choi, M.\ 2005, \apj, 626, 919 

\bibitem[Fazio et al.(2004)]{fazio04} Fazio, G.~G., et al.\ 
2004, \apjs, 154, 10 

\bibitem[Flaherty et al.(2007)]{flaherty07} Flaherty, K.~M., 
Pipher, J.~L., Megeath, S.~T., Winston, E.~M., Gutermuth, R.~A., Muzerolle, 
J., Allen, L.~E., \& Fazio, G.~G.\ 2007, \apj, 663, 1069 

\bibitem[Frerking \& Langer(1982)]{frerking82} Frerking, M.~A., \&
Langer, W.~D.\ 1982, \apj, 256, 523

\bibitem[G{\^a}lfalk \& Olofsson(2007)]{galfalk07} G{\^a}lfalk, M., \&
Olofsson, G.\ 2007, \aap, 475, 281

\bibitem[Galv{\'a}n-Madrid et al.(2004)]{galvan04} Galv{\'a}n-Madrid,
R., Avila, R., \& Rodr{\'{\i}}guez, L.~F.\ 2004, Revista Mexicana de
Astronomia y Astrofisica, 40, 31

\bibitem[Gee et al.(1985)]{gee85} Gee, G., Griffin,
M.~J., Cunningham, T., Emerson, J.~P., Ade, P.~A.~R., \& Caroff,
L.~J.\ 1985, \mnras, 215, 15P

\bibitem[Gordon et al.(2005)]{gordon05} Gordon, K.~D., et al.\ 
2005, \pasp, 117, 503 

\bibitem[Harvey et al.(2001)]{harvey01} Harvey, D.~W.~A., Wilner,
D.~J., Lada, C.~J., Myers, P.~C., Alves, J.~F., \& Chen, H.\ 2001,
\apj, 563, 903

\bibitem[Harvey et al.(2003a)]{harvey03a} Harvey, D.~W.~A., Wilner,
D.~J., Myers, P.~C., \& Tafalla, M.\ 2003a, \apj, 596, 383

\bibitem[Harvey et al.(2003b)]{harvey03b} Harvey, D.~W.~A., Wilner,
D.~J., Myers, P.~C., Tafalla, M., \& Mardones, D.\ 2003b, \apj, 583,
809

\bibitem[Hirano et al.(1988)]{hirano88} Hirano, N., Kameya, O., 
Nakayama, M., \& Takakubo, K.\ 1988, \apjl, 327, L69 

\bibitem[Hodapp(1998)]{hodapp98} Hodapp, K.-W.\ 1998, \apjl, 500, L183

\bibitem[Houck et al.(2004)]{houck04} Houck, J.~R., et al.\ 
2004, \apjs, 154, 18

\bibitem[Huard et al.(1999)]{huard99} Huard, T.~L., Sandell, G., \&
Weintraub, D.~A.\ 1999, \apj, 526, 833

\bibitem[Indebetouw et al.(2005)]{indebetouw05} Indebetouw, R., et 
al.\ 2005, \apj, 619, 931

\bibitem[J{\o}rgensen et al.(2007)]{jorgensen07} J{\o}rgensen, J.~K.,
et al.\ 2007, \apj, 659, 479

\bibitem[Keene et al.(1980)]{keene80} Keene, J., Hildebrand, R.~H.,
Whitcomb, S.~E., \& Harper, D.~A.\ 1980, \apjl, 240, L43

\bibitem[Keene(1981)]{keene81} Keene, J.\ 1981, \apj, 245, 115

\bibitem[Keene et al.(1983)]{keene83} Keene, J., Davidson, J.~A.,
Harper, D.~A., Hildebrand, R.~H., Jaffe, D.~T., Loewenstein, R.~F.,
Low, F.~J., \& Pernic, R.\ 1983, \apjl, 274, L43

\bibitem[Krumholz et al.(2007)]{krumholz07} Krumholz, M.~R., Klein,
R.~I., \& McKee, C.~F.\ 2007, \apj, 665, 478

\bibitem[Kutner \& Ulich(1981)]{kutner81} Kutner, M.~L., \& Ulich,
B.~L.\ 1981, \apj, 250, 341

\bibitem[Lallement et al.(2003)]{lallement03} Lallement, R., Welsh,
B.~Y., Vergely, J.~L., Crifo, F., \& Sfeir, D.\ 2003, \aap, 411, 447

\bibitem[Looney et al.(2007)]{looney07} Looney, L.~W., Tobin, J.~J.,
\& Kwon, W.\ 2007, \apjl, 670, L131

\bibitem[Lucy(1974)]{lucy74} Lucy, L.~B.\ 1974, \aj, 79, 745 

\bibitem[Makovoz \& Marleau(2005)]{makovoz05} Makovoz, D., \& Marleau,
F.~R.\ 2005, \pasp, 117, 1113

\bibitem[Martin \& Barrett(1978)]{martin78} Martin, R.~N., \& Barrett,
A.~H.\ 1978, \apjs, 36, 1

\bibitem[Mozurkewich et al.(1986)]{mozu86} Mozurkewich,
D., Schwartz, P.~R., \& Smith, H.~A.\ 1986, \apj, 311, 371 


\bibitem[Nelson \& Langer(1997)]{nelson97} Nelson, R.~P., \& Langer,
W.~D.\ 1997, \apj, 482, 796

\bibitem[Noriega-Crespo et al.(2004)]{noriega04} Noriega-Crespo, 
A., et al.\ 2004, \apjs, 154, 352 

\bibitem[Ossenkopf \& Henning(1994)]{ossenkopf94} Ossenkopf,
V., \& Henning, T.\ 1994, \aap, 291, 943

\bibitem[Richardson(1972)]{richardson72} Richardson, W.~H.\ 1972,
Journal of the Optical Society of America (1917-1983), 62, 55

\bibitem[Rieke \& Lebofsky(1985)]{rieke85} Rieke, G.~H., \& Lebofsky,
M.~J.\ 1985, \apj, 288, 618

\bibitem[Rieke et al.(2004)]{rieke04} Rieke, G.~H., et al.\ 
2004, \apjs, 154, 25 

\bibitem[Robitaille et al.(2006)]{robitaille06} Robitaille, T.~P., 
Whitney, B.~A., Indebetouw, R., Wood, K., 
\& Denzmore, P.\ 2006, \apjs, 167, 256 

\bibitem[Robitaille et al.(2007)]{robitaille07} Robitaille, T.~P., 
Whitney, B.~A., Indebetouw, R., \& Wood, K.\ 2007, \apjs, 169, 328 

\bibitem[Saito et al.(1999)]{saito99} Saito, M., Sunada, K., Kawabe,
R., Kitamura, Y., \& Hirano, N.\ 1999, \apj, 518, 334

\bibitem[Sault et al.(1995)]{sault95} Sault R.J., Teuben P.J., Wright
M.C.H., 1995, in Astronomical Data Analysis Software and Systems IV,
ed. R. Shaw, H.E. Payne, J.J.E. Hayes, ASP Conf. Ser., 77, 433-436

\bibitem[Shirley et al.(2000)]{shirley00} Shirley, Y.~L., Evans,
N.~J., II, Rawlings, J.~M.~C., \& Gregersen, E.~M.\ 2000, \apjs, 131,
249

\bibitem[Shirley et al.(2002)]{shirley02} Shirley, Y.~L., Evans, 
N.~J., II, \& Rawlings, J.~M.~C.\ 2002, \apj, 575, 337 

\bibitem[Smith et al.(2007)]{smith07} Smith, J.~D.~T., et al.\ 
2007, \pasp, 119, 1133 

\bibitem[Stutz et al.(2007)]{stutz07} Stutz, A.~M., et al.\ 2007,
\apj, 665, 466

\bibitem[Tomita et al.(1979)]{tomita79} Tomita, Y., Saito, T., \&
Ohtani, H.\ 1979, \pasj, 31, 407

\bibitem[van Dishoeck(2004)]{vandishoeck04} van Dishoeck, E.~F.\ 2004,
\araa, 42, 119

\bibitem[Velusamy et al.(1995)]{velusamy95} Velusamy, T., Kuiper,
T.~B.~H., \& Langer, W.~D.\ 1995, \apjl, 451, L75

\bibitem[Velusamy et al.(2007)]{velusamy07} Velusamy, T., Langer, 
W.~D., \& Marsh, K.~A.\ 2007, \apjl, 668, L159 

\bibitem[Velusamy et al.(2008)]{velusamy08} Velusamy, T., Marsh, K.A.,
Beichman, C.A., Backus, C.R., Thompson, T.J.2008, AJ (in press: to
appear in July issue)

\bibitem[Walker et al.(1990)]{walker90} Walker, C.~K., Adams, 
F.~C., \& Lada, C.~J.\ 1990, \apj, 349, 515 

\bibitem[Whitehouse \& Bate(2006)]{whitehouse06} Whitehouse, S.~C., \&
Bate, M.~R.\ 2006, \mnras, 367, 32

\bibitem[Wilner et al.(2000)]{wilner00} Wilner, D.~J., Myers, P.~C.,
Mardones, D., \& Tafalla, M.\ 2000, \apjl, 544, L69

\bibitem[Wu et al.(2007)]{wu07} Wu, J., Dunham, M.~M., 
Evans, N.~J., II, Bourke, T.~L., \& Young, C.~H.\ 2007, \aj, 133, 1560 

\bibitem[Zhou et al.(1990)]{zhou90} Zhou, S., Evans, N.~J., II,
Butner, H.~M., Kutner, M.~L., Leung, C.~M., \& Mundy, L.~G.\ 1990,
\apj, 363, 168

\end{thebibliography}
\end{document}